\documentclass[sigplan,nonacm]{acmart}

\usepackage[compact]{titlesec} 
\usepackage{tikz}
\usepackage{amsmath,pifont} 
\usepackage[english]{babel} 

\usepackage{algorithm}
\usepackage{algorithmicx}
\usepackage[noend]{algpseudocode}
\usepackage{float}

\newcommand{\eg}{{\it e.g.,\ }}
\newcommand{\etc}{{\it etc.\ }}

\newcommand{\ie}{{\it i.e.,\ }}

\newcommand{\us}{\,$\mu$s\xspace}
\newcommand{\ms}{\,ms\xspace}

\usepackage[]{hyperref}
\usepackage[capitalise]{cleveref}
\crefname{section}{}{\S\S}
\crefdefaultlabelformat{\S#2#1#3}
 
\makeatletter
\def\BState{\State\hskip-\ALG@thistlm}
\renewcommand\theHALG@line{\thealgorithm.\arabic{ALG@line}}
\makeatother

\usepackage{utfsym}
\usepackage{xspace}
\usepackage{comment}
\usepackage[normalem]{ulem}
\usepackage{enumitem}
\setlist[enumerate]{itemsep=0mm}
\usepackage{multirow}
\usepackage{float}
\usepackage{makecell}

\setlist{nolistsep}
\usepackage{url}

\usepackage{breakurl}

\usepackage{listings}
\usepackage{color}
\usepackage[font=small]{caption}
\usepackage[font=small]{subcaption}
\usepackage{tabularx}

\usepackage{textcomp}
\usepackage{booktabs}

\definecolor{codegreen}{rgb}{0,0.6,0}
\definecolor{codegray}{rgb}{0.5,0.5,0.5}
\definecolor{codepurple}{rgb}{0.58,0,0.82}
\definecolor{backcolour}{rgb}{0.95,0.95,0.92}

\lstdefinestyle{mystyle}{
  language=C,
  float=tp,
  floatplacement=tbp,
  backgroundcolor=\color{backcolour},   commentstyle=\color{codegreen},
  keywordstyle=\color{magenta},
  numberstyle=\tiny\color{codegray},
  stringstyle=\color{codepurple},
  basicstyle=\ttfamily\footnotesize,
  breakatwhitespace=false,         
  breaklines=true,                 
  captionpos=b,                    
  keepspaces=true,                 
  numbers=left,                    
  numbersep=5pt,                  
  showspaces=false,                
  showstringspaces=false,
  showtabs=false,                  
  tabsize=2,
  xleftmargin=1.5em, 
}

\usepackage{balance}

\newcommand{\sysn}{Meili} 
\newcommand{\sys}{Meili\xspace} 
\newcommand{\ca}{CA\xspace}
\newcommand{\tor}{TO\xspace}
\newcommand{\tors}{TOs\xspace}
\newcommand{\bluef}{BlueField-2\xspace}
\newcommand{\bfone}{BF-1\xspace}
\newcommand{\bftwo}{BF-2\xspace}
\newcommand{\snic}{SmartNIC\xspace}
\newcommand{\snics}{SmartNICs\xspace}
\newcommand{\saas}{SNICaaS\xspace}
\newcommand{\nic}{NIC\xspace}
\newcommand{\nics}{NICs\xspace}

\newcommand{\se}{SE\xspace}
\newcommand{\uco}{UCF\xspace}
\newcommand{\ucos}{UCFs\xspace}
\newcommand{\ucf}{UCF\xspace}
\newcommand{\ucfs}{UCFs\xspace}

\graphicspath{{figures/}}

\renewcommand\footnotetextcopyrightpermission[1]{} 
\setcopyright{none}
\hypersetup{final}
\begin{document}
\sloppy

\pagestyle{plain}

\title{\sys: Enabling \snic as a Service in the Cloud}

\author{Qiang Su}
\affiliation{
\institution{CUHK}
\city{Hong Kong SAR}
\country{China}
}
\author{Shaofeng Wu}
\affiliation{
\institution{CUHK}
\city{Hong Kong SAR}
\country{China}
}
\author{Zhixiong Niu}
\affiliation{
\institution{Microsoft Research}
\city{Beijing}
\country{China}
}
\author{Ran Shu}
\affiliation{
\institution{Microsoft Research}
\city{Beijing}
\country{China}
}
\author{Peng Cheng}
\affiliation{
\institution{Microsoft Research}
\city{Beijing}
\country{China}
}
\author{Yongqiang Xiong}
\affiliation{
\institution{Microsoft Research}
\city{Beijing}
\country{China}
}
\author{Zaoxing Liu}
\affiliation{
\institution{University of Maryland}
\city{College Park}
\country{USA}
}
\author{Hong Xu}
\affiliation{
\institution{CUHK}
\city{Hong Kong SAR}
\country{China}
}

\begin{abstract}

\snics are touted as an attractive substrate for network application offloading, offering benefits in programmability and host resource saving.
The current usage restricts offloading to local hosts and confines \snic ownership to individual application teams, resulting in poor resource efficiency and scalability. 
This paper presents \sys, a novel system that realizes \snic as a service to address these issues. 
\sys organizes heterogeneous \snic resources as a pool and offers a unified one-NIC abstraction to application developers.
This allows developers to focus solely on the application logic while operators efficiently optimize resource allocation for performance needs.
Our evaluation on NVIDIA BlueField series and AMD Pensando \snics demonstrates that \sys improves cluster resource efficiency up to 1.75$\times$ compared to common approaches in state-of-the-art systems, and achieves scalable throughput with low latency overhead.

\end{abstract}
 
\maketitle
\pagestyle{plain}

\section{Introduction}
\label{sec:introduction}

Modern data centers have witnessed a notable gap between the stagnation of CPU power and the increase in network bandwidth,  
promoting the offloading of certain tasks to network hardware as a solution \cite{FPMC18,clara21,e319,floem18,fairnic20,TOE,pfaff2015ovs,GVPG14}. 
\snics, equipped with SoC cores and diverse domain-specific hardware accelerators, 
have gained widespread deployment in production data centers for offloading \cite{mbf1,mbf2,stringray,ntn,pensando}.

The current use of \snics in the cloud faces three prominent challenges.
(1) First, individual \snics have wimpy and limited onboard resources \cite{pensando,stringray,ntn,liquid,mbf1,mbf2}, and struggle to meet diverse resource demands from applications.  
While more resourceful \nics are certainly helpful, the pace of hardware development lags behind the rapid evolution of software and services.
Some prior work mitigates this by migrating onboard processing to the host at runtime \cite{le17uno}, which reduces the potential savings on host resources.
Other efforts attempt to enhance resource elasticity by disaggregating programmable ASICs \cite{ns21,db23,dyk23}, but deploying the entire application program on the monolithic \snic and scaling onboard resources at per-\nic granularity, leading to suboptimal resource efficiency.
Moreover, they limit the offloading to P4-compatible hardware.
(2) Second, despite existing efforts on co-running onboard applications \cite{panic20,supernic22,nica19,fairnic20}, 
sharing \snics today is inefficient as they are owned by individual application teams.
This requires resource coordination between teams, which is time-consuming and error-prone. 
It also leads to redundant management labor, slowing down application development and deployment.
(3) Third, due to lack of global visibility into the \snic cluster, operators cannot perform essential management tasks effectively like adaptive resource scaling.

\begin{figure}[t]
    \centering
    \includegraphics[width=\linewidth]{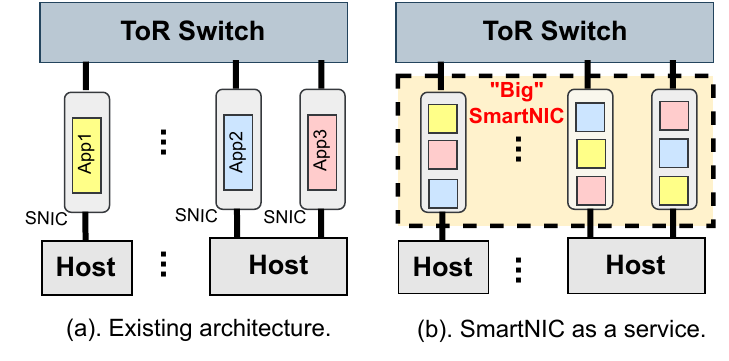}
    \vspace{-7mm}
    \caption{Our vision: \snic as a service.}
    \label{fig:idea_overview}
\end{figure}

We thus envision a unified platform to provide \snic as a service (\saas) as shown in Figure~\ref{fig:idea_overview}. 
Similar to platform as a service (PaaS) in cloud computing, \saas offers efficient resource management for a pool of \snics to reduce the development and deployment cost. 
Developers only need to create applications to a big \snic with all resources ever needed and submit them for deployment with performance targets. 
With \saas, operators gain fine-grained control over \snic resources in the cluster, facilitating tasks like allocation, \nic multiplexing, and adaptive scaling. 
For example, when deploying three applications as shown later in \S\ref{subsec:benefit}, 
existing \snic platforms require separate deployment in three \nics. 
Many resources on each \nic are underutilized this way as different applications have different demands. 
In contrast, with \saas, only two \nics are actually needed to deliver the same performance.

To realize this vision, we build a new system \sys for the popular SoC based \snics. 
In a nutshell, \sys unifies heterogeneous \snics as one big resource pool and exposes a simple ``one-\nic'' abstraction for applications to tap into this pool efficiently and flexibly. An application pipeline is potentially deployed across different \nics, and a \nic potentially hosts different applications. 
In \sys, we need to address three imminent design challenges. 

{\bf Programming model.}  
The current programming model that assumes a monolithic \snic as the target cannot deploy a program to a pool of distributed and heterogeneous compute resources \cite{floem18,ipipe19,clara21,fairnic20}. 
Deploying an application entirely on a single accelerator that only provides fixed functionalities (\eg crypto, regex) is also typically impractical. 
Therefore, we design a new modular programming model that composes an application with small offloading modules 
so that they can be deployed across the resource pool to harness their collective benefits.
Our model provides common packet- and socket-level processing operations out-of-the-box; users can also write their own functions, \ie user-customized functions (\ucfs). 
Each function may run on heterogeneous resources for maximal performance and efficiency. 
We also develop a state API for applications to manage their states as if they are running locally, with the heavy-lifting taken care of by a state engine.

{\bf Scalable data plane.} 
A \sys application is naturally executed in a pipeline fashion with each function as an independent stage. 
To achieve scalable throughput, we replicate the execution pipeline to additional resources and load balance the ingress traffic accordingly. 
Yet, unlike conventional pipeline deployments on homogeneous targets \cite{nfvnice17,pd19,gp19},  
the pipeline stages in \sys exploit heterogeneous resources with vastly different processing latencies.
Replicating a full pipeline does not reduce the pipeline bubbles or resource underutilization. 
\sys thus performs partial pipeline replication and scales each pipeline stage independently. The key idea here is to replicate only the bottleneck stages to the point such that the shortest stage becomes fully utilized with no bubbles.  
The traffic is partitioned across parallel stage replicas on a flow basis.
Another issue with parallel replicas is that they necessitate concurrent access to the shared traffic buffer, and may inflate latency due to buffer locking.
To mitigate this, \sys allocates a dedicated lockless ring buffer for each replica and distributes traffic among them.

{\bf Unified control plane.}
\sys needs a central controller to orchestrate cluster-wise resource allocation while meeting application performance targets.
Although distributing pipeline stages across \snics improves efficiency,
the inter-stage traffic redirection over network introduces overhead on \nic and link bandwidth, and end-to-end latency.
Therefore, we propose a locality-aware resource allocation approach that prioritizes placing consecutive stages on the same \nic.
Our approach selects the \nics for each stage based on its preceding stage's location, favoring the highest onboard compute resources {when the preceding stage resides in multiple \nics}. 
\sys controller periodically synchronizes resource and application status with each \nic, and conducts adaptive scaling as needed.

We implement a prototype of \sys in C, which supports NVIDIA BlueField \cite{mbf1,mbf2} and AMD Pensando \snics \cite{pensando}. 
Our code is open source anonymously at \cite{opensource}. 
Our testbed evaluation results show that \sys outperforms two typical resource allocation approaches in state-of-the-art \snic systems \cite{dyk23,db23,e319,le17uno} by 1.75$\times$ and 1.48$\times$, respectively, in overall resource efficiency.
\sys also delivers scalable throughput with multiple pipelines with small latency overheads,  facilitates better resource availability in the pool, and enhances host CPU saving.

\section{Motivation}
\label{sec:motivation}

We start by discussing the motivation and opportunity for \snic as a service.

\subsection{Current Practices and Limitations}
\label{subsec:use}

The current \snic usage in the cloud has interesting characteristics.
First, it is mostly used to serve first-party services, particularly those involving virtualization and network functions such as stateful firewall, network intrusion detection, and IPSec. 
These services are developed and maintained by cloud providers \cite{FPMC18,clicknp16,clara21,db23}. 
Moreover, individual application teams can directly write, configure, and deploy applications to the \snics with maximum flexibility.\footnote{From a private discussion with a leading commercial cloud provider.} 
Second, as shown in Figure~\ref{fig:idea_overview}(a) before, the offloading is local in scope and applications are server-centric, 
meaning that a \snic only serves traffic of the host it is physically connected to, and operates directly on the data-path of the host applications. 
These practices are becoming a roadblock to efficient utilization of onboard resources, 
given the continuing popularity of offloading to \snics.

\begin{figure}[t]
    \centering
    \includegraphics[width=\linewidth]{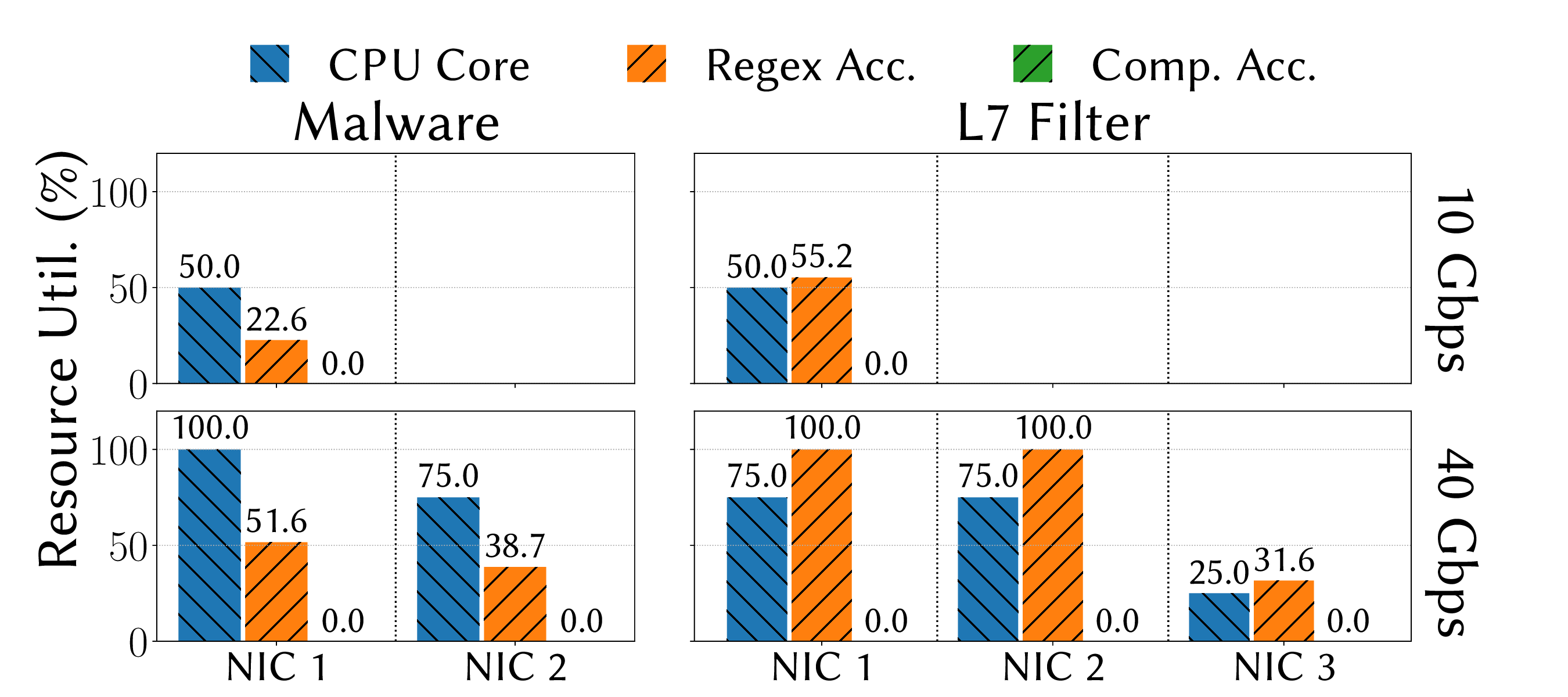}
    \vspace{-7mm}
    \caption{The resource utilization on each \snic to meet the throughput targets for Malware Detection and L7 Filter.
    CPU utilization is the ratio of used cores to the total number of cores; 
    Accelerator utilization is calculated through dividing the end-to-end application throughput by its full capacity. }
    \label{fig:reuti}
\end{figure}

\noindent{\bf Inefficient resource scaling.}
Despite wide adoption, \snic's wimpy onboard resources pose challenges to meet applications' diverse and dynamic resource demands.
Current solutions address this by replicating the entire application to additional \nics, scaling at the per-\nic granularity due to local-only offloading, and load balancing traffic across them \cite{dyk23,db23,e319}. 
This approach, however, results in suboptimal resource efficiency.
We deploy two common applications, Malware Detection \cite{snortips} and L7 Filter \cite{l7filter,dpinv}, 
on \bluef \snics with the MACCDC traffic trace \cite{maccdc} and open-source Snort rules \cite{snt3rule}. 
Figure~\ref{fig:reuti} shows the utilization of onboard resources to achieve varying throughput targets.  
For instance, to reach 40~Gbps for L7 Filter, three \nics are needed and \nics 1 and 2's regex accelerator are fully utilized because L7 Filter is bound by the regex accelerator.  
On the other hand, CPU on each \nic is 75\% utilized at best, and the compression accelerators are completely idle.
Similar inefficiency is observed in scaling Malware Detection, for which CPU is the bottleneck.
With two \nics it reaches 40~Gbps, but the regex and compression accelerators are underutilized.

\begin{figure}[t]
    \centering
    \includegraphics[width=\linewidth]{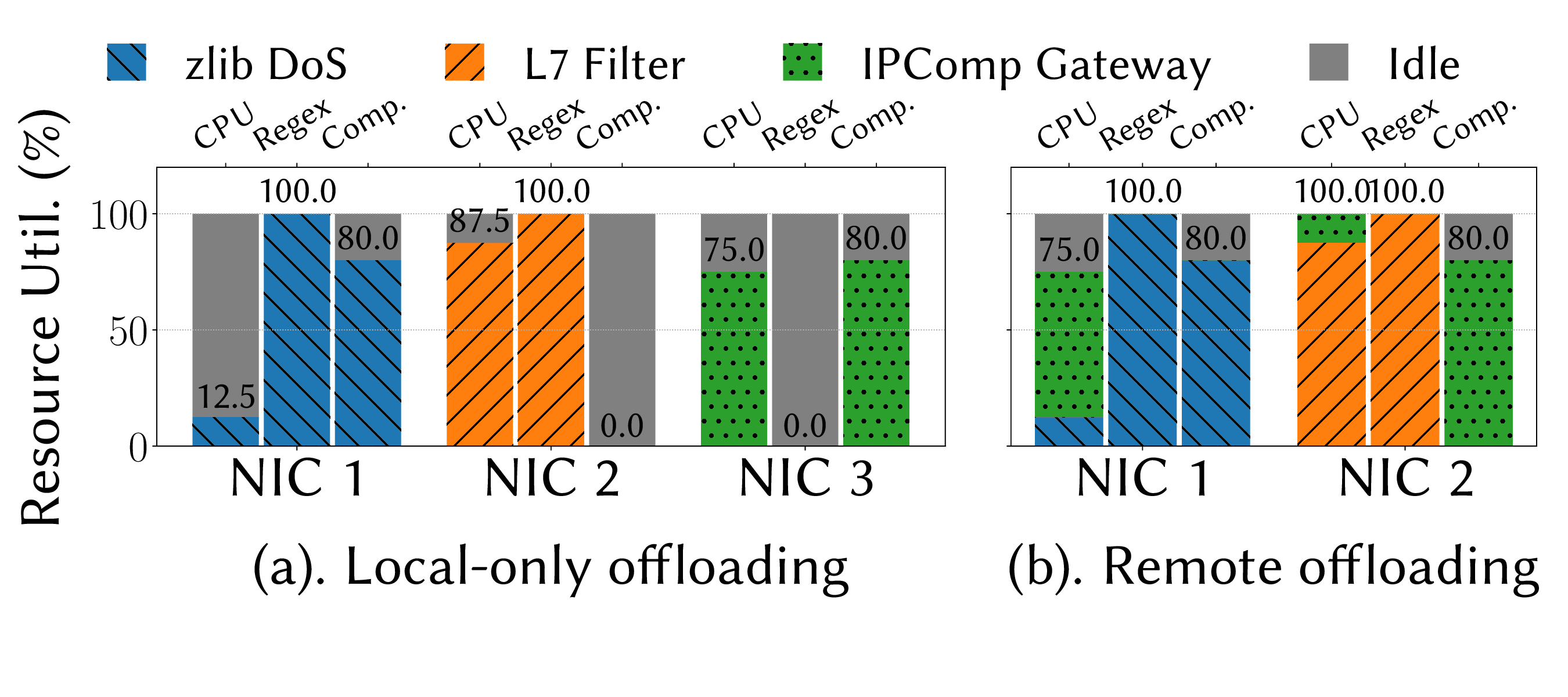}
    \vspace{-13mm}
    \caption{The resource utilization of deploying three applications for 20Gbps throughput by multiplexing \texttt{NIC1} and \texttt{NIC2}. 
    In (b), IPComp Gateway runs on both \nics.
    }
    \label{fig:muti_on}
\end{figure}

\begin{table}[t]
    \centering
    \footnotesize
    \resizebox{.9\linewidth}{!}{
    \begin{tabular}[t]{ccccc}
    \toprule
    App. & Deployment & Avg lat. & 90\%ile lat. & 99\%ile lat. \\
    \midrule
    \multirow{2}{*}{zlib DoS} & Individual & 6.32 & 6.62 & 7.01\\
                           & Multiplex & 6.36 & 6.63 & 7.08 \\ 
    \midrule
    \multirow{2}{*}{L7 Filter} & Individual & 5.85 & 5.93 & 6.17 \\    
                           & Multiplex & 5.83 & 5.91 & 6.37 \\ 
    \midrule
    \multirow{2}{*}{IPComp Gateway} & Individual & 5.49 & 5.67 & 5.77 \\    
                           & Multiplex & \textbf{9.24} & \textbf{9.57} & \textbf{10.11} \\ 
    \bottomrule
    \end{tabular}
    }
    \captionof{table}{Application latencies (\us) of multiplexing two \snics.}
    \label{table:colat}
\end{table}

\noindent{\bf Inflexible orchestration.}
Managing \snics by individual application teams also impedes global resource orchestration.
Each application has a strong tendency to overbook resources for failover and future demands.
This creates additional contention among applications and a substantial decline in utilization.
More fundamentally, essential management functions such as isolation, monitoring, and troubleshooting are also independently handled, leading to redundant efforts and suboptimal decisions due to the lack of the global view. 
To alleviate these difficulties, teams now have to coordinate manually and on a case-by-case basis, which is time-consuming and error-prone.

\subsection{What Can \sys Bring?}
\label{subsec:benefit}

To address these issues, we advocate a unified platform that manages \snics as one resource pool, enabling flexible and efficient resource management while meeting dynamic application performance targets.
We showcase the potential benefits of such a platform.

\noindent{\bf Cluster-wise resource efficiency via fine-grained multiplexing.}
Unlike prior work that allocates resources at the \nic granularity \cite{dyk23,db23,e319}, 
pooling \snic resources allows for finer-grained allocation and higher efficiency. 
To illustrate this, we conduct experiments on three BlueField-2 \snics, 
deploying three applications --- zlib DoS \cite{zdos}, L7 Filter \cite{l7filter}, and IPComp Gateway \cite{ipcomprfc,nfp17} with varying resource requirements.
We generate 1500B packets via DPDK-Pktgen, replay traffic traces from MACCDC \cite{maccdc}, and deploy Docker containers to run applications.
Figure~\ref{fig:muti_on}(a) depicts that each \nic has vacant resources under local-only offloading when the throughput target is 20Gbps. 
By pooling, we can consolidate applications onto just two \nics: IPComp Gateway now runs on NIC1's CPU and NIC2's compression accelerator with traffic re-direction using DPDK, thus enabling remote offloading. 
Figure~\ref{fig:muti_on}(b) displays the elevated utilization after multiplexing. 
The three apps are deployed one-by-one and all reach the throughput target.
Latency-wise, zlib DoS and L7 Filter are unaffected when co-running with IPComp Gateway, while the latter experiences increased latency as in Table~\ref{table:colat} due to distributed deployment.

\begin{figure}[t]
    \centering
    \includegraphics[width=\linewidth]{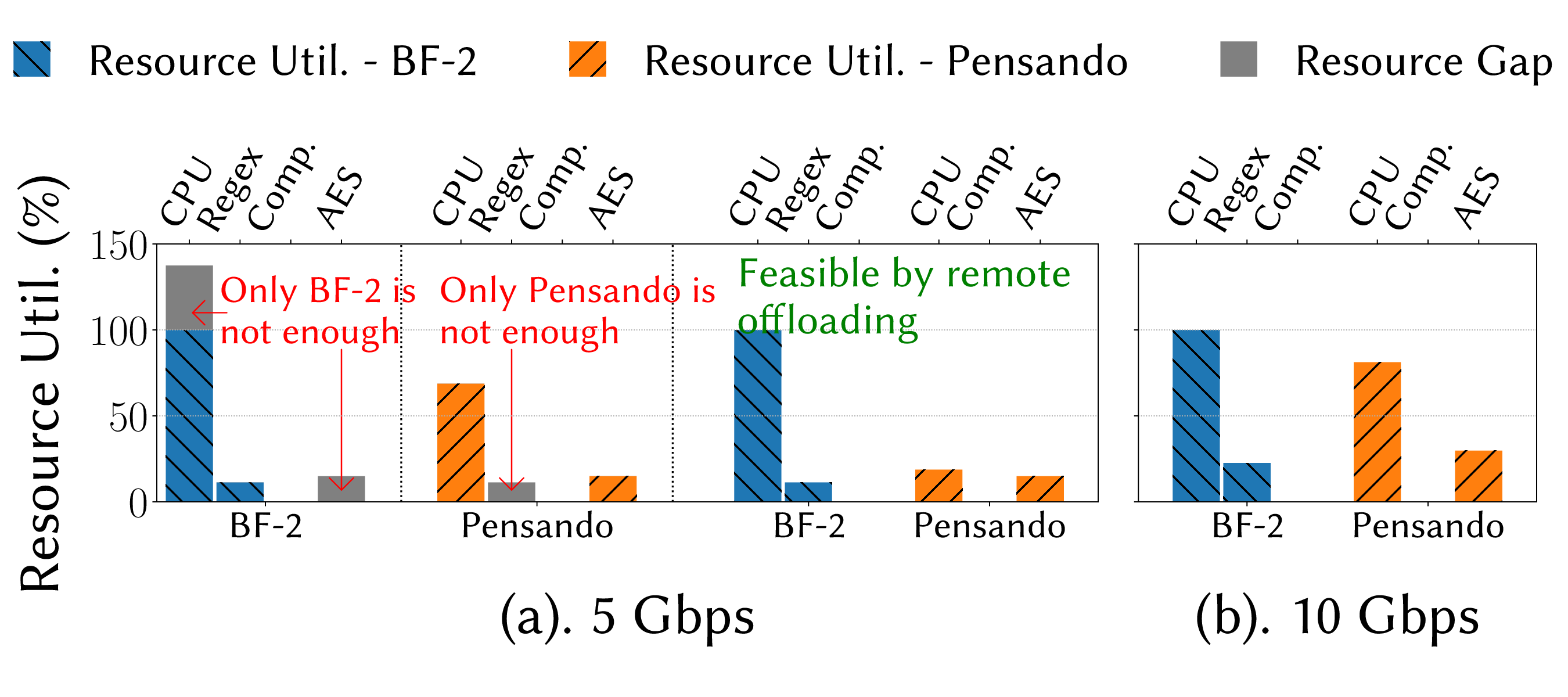}
    \vspace{-10mm}
    \caption{IPsec Gateway deployment on BlueField-2 and Pensando.}
    \label{fig:moti_ipsec}
\end{figure}

This example also presents a new tradeoff between overall cluster-wise resource utilization and application latency overheads. 
Therefore, it is important to deploy latency-sensitive applications with the conventional local-only offloading 
while distributing others across the cluster with remote-offloading to optimize overall resource efficiency. 

\noindent{\bf Better resource availability.}
With remote-offloading and resource pooling, an application can tap into heterogeneous resources from different \snics.
To understand this, we showcase an IPsec Gateway \cite{e319,wl1,dyk23,nfp17} that utilizes CPU, regex, and AES accelerators.
Specifically, we employ a BlueField-2 and a Pensando \snic, with a target throughput of 5Gbps. 
As shown in Figure~\ref{fig:moti_ipsec}(a), deploying this application on either \nic is infeasible (BlueField-2 misses AES and Pensando misses regex accelerators). 
However, by pooling the two together, we can deploy the application successfully by using the CPU and regex accelerator on BlueField-2 and the CPU and AES accelerator on Pensando. 

\noindent{\bf Adaptive resource scaling.}
With ample resources in the pool, it is also possible for the platform to adaptively adjust resource allocation and application placement to meet the evolving performance targets in real-time. 
For example, developers can submit a new 10Gbps throughput requirement for IPsec Gateway to the platform.
As shown in Figure~\ref{fig:moti_ipsec}(b), the platform may allocate additional resources from the two \nics (mostly from Pensendo) to achieve it.

\subsection{Scope and Applicability}
In many cases, \snic applications  work independently from the host logic. 
Common examples include network functions like firewalls, L7 filters, API gateways, among others. 
This decoupling allows us to run them across multiple \snics.
However, there are cases where a host application offloads a specific part of its logic, and the offloaded part has frequent interaction with the host. 
Examples include offloading the TCP/IP stack \cite{flextoe22,iotcp23} and network-attached storage \cite{xenic21,linefs21,gimbal21,kvdirect17}.
These frameworks impose high barriers to distributing the onboard logic across the cluster.
Therefore, our vision is not suitable for these use cases; their current deployment can remain unchanged.
Moreover, \sys manages \snics \textit{within each rack} as a separate pool,
as pooling across racks can be vulnerable to common network issues such as congestion.

\noindent{\bf Why not microservices?}
One might be wondering if existing microservice platforms \cite{sf19,ganmic,k8s,istio} can be applied to deploy and scale different parts of a \snic application independently as a solution. 
Unfortunately, the high communication overheads using remote procedure call (RPC) alone very likely outweigh the potential elasticity benefits of this approach, (\eg over 40\% latency increase on x86 CPU as profiled in \cite{ganmic,nikita21dagger}). 
Note that common \nic applications usually favor simpler traffic redirection \cite{FPMC18,clicknp16,clara21,db23}.
These platforms are designed for Web services and also incur major overheads for managing components such as databases which are not necessary for \snics. 
On the other hand, essential aspects such as the unified programming model, efficient data plane design for high performance, \etc are all missing in these platforms.

\section{Overview}
\label{sec:overview}

\begin{figure}[t]
    \vspace{-3mm}
    \centering
    \includegraphics[width=\linewidth]{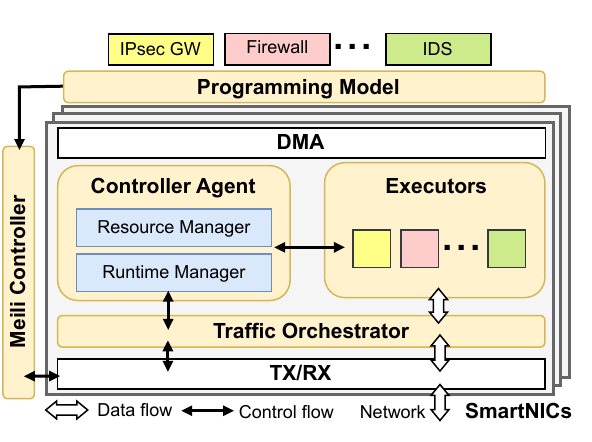}
    \vspace{-8mm}
    \caption{Overview of \sys's architecture.}
    \vspace{-3mm}
    \label{fig:overview}
\end{figure}

\noindent{\bf Architecture.}
\sys is a unified platform for heterogeneous resources including programmable SoC cores and hardware accelerators.
Figure~\ref{fig:overview} illustrates its architecture.
It has three main components: 
a programming model, 
a control plane featuring a \sys Controller and per-\snic Controller Agents (\ca), 
and a data plane with per-\snic Traffic Orchestrators (\tor) and the application runtime Executor.

\noindent{\underline{\it Programming model.}}
\sys offers a modular Programming Model that offers a one-NIC abstraction to facilitate the development and deployment of \snic applications.
Developers build each application (\eg IPsec Gateway and Firewall) as multiple modules using this model and submit it, along with performance requirements, to \sys Controller.

\noindent{\underline{\it Control plane.}}
The \sys Controller orchestrates the global placement of applications to meet their performance demands while optimizing cluster-wise resource allocation.
Each \ca employs a Resource Manager and a Runtime Manager to monitor onboard resource usage (\eg CPU and hardware accelerators) and application status (\eg latency), synchronizing them with \sys Controller periodically. Upon receiving application programs and performance targets, \sys Controller determines resource demands through offline profiling, generates global resource allocation and application placement policies, and sends them to corresponding {\ca}s, which then configure the resource allocation locally.

\noindent{\underline{\it Data plane.}}
Each application operates within a pipeline paradigm, where each module runs as a stage and inside an isolated runtime called Executor (\eg Container).
The data plane also enforces resource-efficient pipeline parallelism while achieving scalable performance.
A pipeline is usually distributed or replicated across \snics under the same rack.
Additionally, \tor dynamically manages the application traffic to these distributed pipelines (\eg traffic redirection). It also differentiates control traffic from the network and dispatches them to \ca.

\section{Programming Model}
\label{sec:api}

We describe how applications are crafted in \sys to enable deployment across fine-grained resources. 
To guide our design, we analyze the operation patterns of common \snic applications.
First, they are primarily designed for network traffic processing and work in a streaming pattern. 
Second, they generally operate on two abstractions:
1) {packet}, where application logic is on the packet or flow level (\eg IPSec and Firewall);
2) {socket}, where application logic relies on semantics above the transport layer (\eg API gateway and L7 load balancer).

\noindent{\bf Our approach.}
In \sys each application is composed as a directed acyclic graph of functions which are independently deployed on different \nics and resources.
Based on our observations above, \sys offers \textit{packet processing} and \textit{socket processing} paradigms (\cref{subsec:bp}), where the exact processing logic is expressed through a \textit{user-customized function} (\uco). 
\sys provides accelerator function APIs for hardware-agnostic invocation of accelerators (\cref{subsec:afa}). 
It also exposes state management APIs for stateful applications (\cref{subsec:state}).
\Cref{subsec:meili_api_list} presents a partial list of \sys's APIs.

\noindent{\bf An example.}
Listing~\ref{listing:nf_example} is a packet processing example using \sys's APIs.
The user writes her \ucos (\eg \texttt{ddos\_check}, \texttt{url\_check}, and \texttt{ipsec}) 
and compose them using \sys's basic paradigms.
{\texttt{\sys.regex()}} and \texttt{\sys.AES()} are accelerator function APIs for regular expressions and AES encryption, respectively.
We also showcase a socket processing example in \Cref{subsec:sock_ex}.

\subsection{Basic Paradigms}
\label{subsec:bp}

\noindent{\bf Packet Processing.} 
Packet processing typically involves per-packet and per-connection operations. 
As a result, \sys defines two data structures: 1) \textit{\sysn\_packet}, which contains the packet headers, 
the payload, and a reference to the per-packet metadata; 2) \textit{\sysn\_flow}, which contains 
the connection descriptor (\eg 5-tuple) and the per-connection metadata.
Additionally, {\uco}s are callback functions that can access the whole structure and compute the metadata.
\sys provides the following packet processing operations.

\noindent{\underline{\it Packet Transformation --- \texttt{pkt\_trans()}.}}
This takes a packet structure as input and allows users' programs to access, compute, and modify the headers, payload, and per-packet metadata by a \uco, 
such as changing the egress ports. Then it returns the processed packet.

\noindent{\underline{\it Packet Filter --- \texttt{pkt\_flt()}.}}
A packet can be dropped or passed based on the criterion specified in a \uco, with either a \texttt{FLT\_MATCH} or a \texttt{FLT\_UNMATCH} return value. 
Examples include Deep Packet Inspection (DPI) and Firewall based on the payload and per-packet metadata, respectively.

\noindent{\underline{\it Flow Extraction --- \texttt{flow\_ext()}.}}
This operation extracts flows from a stream of packets.
It takes the input as a window size, sliding interval, and a \uco
that defines the rules of constructing flows (\eg five-tuples), and returns the final flow structures.
The packets are by default passed without modification, and can also be processed by 
the above packet transformation and filter operations.

\noindent{\underline{\it Flow Transformation --- \texttt{flow\_trans()}.}}
Similar to the packet transformation, this operates on flow structures 
and allows users to enforce arbitrary {\uco}s on per-connection metadata, 
and returns the processed flow structure at last.

\noindent{\bf Socket Processing.}
\sys's socket processing paradigm follows the typical \texttt{epoll} event mechanism, and supports operations for socket registration and event processing.
Users can register a socket to \sys after a connection is established, allowing \sys to manage the processing on that socket.
Meanwhile, the event processing functionality (\eg \texttt{EPOLL\_IN}) can be crafted as {\uco}s.
The socket processing depends on complete operating systems as it requires TCP/IP stack support (\eg Linux kernel, or user-space stack).

\begin{figure}[t]
\begin{lstlisting}[basicstyle=\tiny\tt,caption={Pseudocode of a packet processing application. It performs  URL filtering \cite{uflt}, IPSec \cite{ipsecbf}, and DDoS detection \cite{ddos}.}, label={listing:nf_example}]
    // User-customized functions
    url_check(Meili_packet pkt) { 
        match_num = Meili.regex(pkt.payload,RULES);
    }
    ipsec(pkt) { 
        encap(pkt);
        sha(pkt, BLK_SIZE);
    }
    ddos_check(Meili_packet pkt) { 
        sum_ent = sum_ent(pkt); 
        joint_ent = joint_ent(pkt);
        if sum_ent - joint_ent > THRESHOLD
            ddos_flag = 1;
    }
    // Meili API invocation
    Meili.pkt_flt(url_check, pkt); 
    Meili.pkt_trans(ipsec, pkt); 
    Meili.AES(pkt, ERY_TAG, BLK_SIZE); 
    Meili.pkt_flt(ddos_check, pkt); 
\end{lstlisting}
\end{figure}

\subsection{Accelerator Function API}
\label{subsec:afa}

\snics integrate various hardware accelerators differing across vendors and generations \cite{mbf2,mbf1,pensando}. 
To abstract this heterogeneity and enable uniform invocation of an accelerator that has diverse implementations in different \nics, \sys introduces a set of Accelerator Function APIs.
Specifically, we abstract shared parameters common to the accelerator, relegating other settings, typically hardware-specific ones, to \sys.
For a compression accelerator, for instance, \sys offers \texttt{\sys.Compress(addr,rt)} on either BlueField-2 \cite{mbf2} or Pensando \snics \cite{pensando}.
Users only need to specify data pointers (\texttt{addr}) and compression ratio (\texttt{rt}), while \sys transmits these values to the accelerator API on specific \snics, 
and configures other parameters like device ID, queue ID, and PCIe addresses. 
Notably, \tor redirects traffic to a remote compression accelerator if the local \nic lacks it.

\subsection{State Management}
\label{subsec:state}

For stateful applications, 
\sys deploys a lightweight per-\nic state engine (\se) and provides a set of state APIs.

\begin{figure}[t]
    \centering
    \vspace{2mm}
    \includegraphics[width=0.85\linewidth]{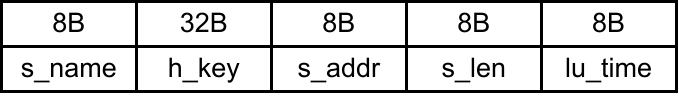}
    \vspace{-2mm}
    \caption{The structure of a state entry. \texttt{s\_name} and \texttt{h\_key} denote the name and hash key of the state; 
\texttt{s\_addr} and \texttt{s\_len} are the memory address and length of the state, which is allocated from dedicated memory regions on each \snic;
and \texttt{lu\_time} denotes the last used time of the state.
It is released if the lifespan \texttt{lu\_time} exceeds a predefined threshold (\eg 500s). }
    \label{fig:state}
\end{figure}

\noindent{\bf State API.}
\sys's state APIs follow the common access pattern: States are writable and readable locally but read-only externally \cite{db23,GVPG14}.
The SE supports six operators: \texttt{ADD}, \texttt{REMOVE}, \texttt{GET}, \texttt{SET}, \texttt{TRAVERSE}, and \texttt{COMPUTE}. 
Users can write customized computation like \texttt{TOP-N} on states in a \ucf and apply it using the \texttt{COMPUTE} operator. 
Figure~\ref{fig:state} illustrates a 64-byte state entry.

\noindent{\bf State engine.}
The \se employs linked hash tables to organize application states, and utilizes RDMA that is compatible with \snic architectures.
When an application calls the state APIs, the \se translates them into corresponding transport operations, such as RDMA WRITE.
For \texttt{GET}, \se first checks local states and, if not found, retrieves the state from remote \snics, by issuing a RDMA READ request to in all \snics running the application. 
However, for \texttt{SET}, \texttt{REMOVE}, and \texttt{ADD}, \se only applies the operations to local states. 
\texttt{TRAVERSE} makes \se fetch linked hash tables from all \snics running the application through RDMA READ and traverse all states locally,
mitigating the communication overhead by reducing RDMA operations.
Similarly, an RDMA WRITE operation transmits \texttt{COMPUTE} instructions to all \snics; then \se performs computation specified in the \ucf and returns aggregated results to the caller.

\section{\sys Data Plane}
\label{sec:dataplane}

To provide scalable performance, we design \sys's data plane with data-path parallelism and traffic orchestration.

\subsection{Data-Path Parallelism}
\label{subsec:dpp}

As a \sys application is composed of chained functions, we formulate its execution in a pipeline pattern with each function as an independent stage (\eg \texttt{ddos\_check} in Listing~\ref{listing:nf_example}).
\sys also deploys an application as multiple pipelines over heterogeneous resources in the rack; the ingress traffic is distributed across pipelines. 

\subsubsection{Resource-Efficient Pipeline Parallelism}
\label{subsub:pipe_para}

\begin{figure}[t]
    \centering
    \includegraphics[width=\linewidth]{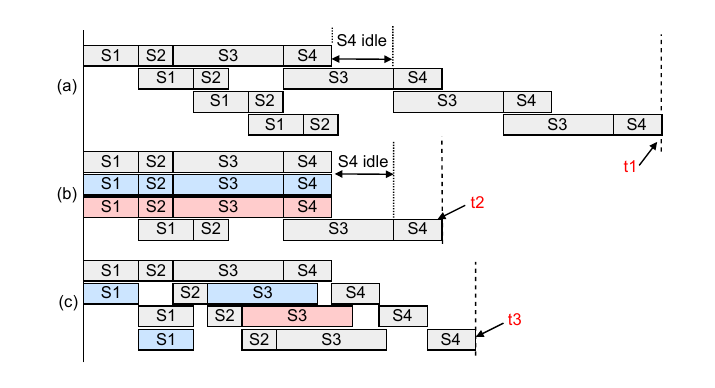}
    \vspace{-8mm}
    \caption{
    The pipeline processing of four flows. 
    (a). The pipeline in Listing~\ref{listing:nf_example}, with stages \texttt{S1} (\texttt{url\_check}), \texttt{S2} (\texttt{encap} and \texttt{sha}), \texttt{S3} (\texttt{\sys.AES}), and \texttt{S4} (\texttt{ddos\_check}). 
    (b). Replicating entire pipeline (3 pipelines).
    (c). Independently scaling individual stages: \texttt{S1} with 2 replications, and \texttt{S3} with 3 replications.
    The stage length represents the per-packet processing latency from offline profiling; colors indicate different replicas for a stage.
    \texttt{t1}, \texttt{t2}, and \texttt{t3} are end-to-end latencies.
    }
    \label{fig:nf_pipe}
\end{figure}

As functions vary in complexity and run on heterogeneous resources, pipeline stages exhibit vastly different processing latency. 
This leads to pipeline bubbles with suboptimal resource utilization and \textit{sub-linear} scaling gain.
For instance, Figure~\ref{fig:nf_pipe}(a) depicts the pipeline of Listing~\ref{listing:nf_example}.
Stages \texttt{S2} and \texttt{S4} underutilize their resources and degrade the overall efficiency. 

\noindent{\bf Strawman solution.}
To scale throughput, the simplest way is to use pipeline parallelism and replicate the entire pipeline. 
As shown in Figure~\ref{fig:nf_pipe}(b), we deploy three pipelines to process the first three packets concurrently.  
However, it does not resolve the inefficiency as the fourth packet still incurs bubbles at \texttt{S2} and \texttt{S4}.
Many prior pipeline deployments on homogeneous targets \cite{nfvnice17,pd19,gp19} attempts to mitigate this issue by dividing tasks into nearly equal-time stages to eliminate bubble,
but this is impractical here: (1) The {\uco}s written by users is the minimum stage granularity, and (2) the built-in functions on hardware accelerators are indivisible black boxes.

\noindent{\bf Our solution.}
Thus we propose \textit{partial pipeline replication} where each stage gets replicated independently of one another. 
By replicating stages with longer processing times more, we can better balance the processing throughput across stages to reduce the bubbles.  
Identifying which stages to replicate and how to replicate then becomes crucial.

Clearly, bubbles occur in a stage when its preceding stages have longer processing times (\eg \texttt{S2} and \texttt{S4} in Figure~\ref{fig:nf_pipe}).
To fully utilize its capacity, its preceding stages need to provide higher throughput. 
We should then replicate its preceding stages to the point where they have the same throughput to it (\eg replicate \texttt{S1} for <\texttt{S1}, \texttt{S2}>).
For the partial pipeline after this bottleneck stage, if there is another bottleneck stage with lower throughput, this partial replication can be applied again, such as <\texttt{S3}, \texttt{S4}> where \texttt{S3} is replicated one more time to reduce bubbles at \texttt{S4}.

We present Algorithm~\ref{alg:pipe_para} based on this intuition to effectively determines the replication factors for each stage for optimal resource efficiency.
It takes pipeline stages {$\mathcal{S}$} and their average per-packet processing latencies {$\boldsymbol{L}$}, determined through offline profiling (\cref{sec:controlplane}), as input.
The algorithm first divides the pipeline into two parts, $\mathcal{S}_{head}$ and $\mathcal{S}_{tail}$, at stage $d$ with the minimal processing latency within $\mathcal{S}$ (line~\ref{alg:a}).
It calculates the respective replication factor for each stage in $\mathcal{S}_{head}$ based on their latency differentials (lines~\ref{alg:c}-\ref{alg:d}).
This process is repeated for the succeeding partial pipeline $\mathcal{S}_{tail}$ (line~\ref{alg:e}) until $\mathcal{S}_{tail}$ only has the last stage.
Figure~\ref{fig:nf_pipe}(c) applies Algorithm~\ref{alg:pipe_para} to the pipeline in Figure~\ref{fig:nf_pipe}(a), with $\boldsymbol{R}=(2\ 1\ 3\ 1)$. Our approach effectively reduces bubbles and improves end-to-end performance compared to (a); unlike (b), it conserves resources on \texttt{S1}, \texttt{S2}, and \texttt{S4}. 
\sys also allocates resources according to $\boldsymbol{R}$ for overall resource efficiency in the control plane (\cref{sec:controlplane}).

\begin{figure}[t]
\small
   \begin{algorithm}[H]
       \centering
       \caption{Partial pipeline replication.}
       \label{alg:pipe_para}
       \begin{algorithmic}[1]
           \State {$\mathcal{S}$: List of pipeline stages}
           \State {$\boldsymbol{L}$: Average per-packet processing latencies of each stage in a pipeline}
           \State {$\boldsymbol{R}$: The number of replications for each stage in a pipeline}
           \Function {num\_replication}{$\mathcal{S}$, $\boldsymbol{L}$}
               \State {$\boldsymbol{R} \leftarrow$ NULL}
               \While {$\mathcal{S} \neq$ NULL} \label{alg:f}
                   \State {$d$ $\leftarrow$ find\_min\_stage($\mathcal{S}$, $\boldsymbol{L}$)} \label{alg:a}
                   \State {$\mathcal{S}_{pre}$, $\mathcal{S}_{post} \leftarrow$ partition($\mathcal{S}$, $d$)} \label{alg:b}
                   \If {$\mathcal{S}_{pre}$ $\neq$ NULL} \label{alg:c}
                       \For {$i$ in $\mathcal{S}_{pre}$}
                           \State {$\boldsymbol{R}_{i}$ $\leftarrow$ $\lceil \boldsymbol{L}_i / \boldsymbol{L}_d\rceil$}\label{alg:d}
                       \EndFor
                   \Else 
                       \State {$\mathcal{S} \leftarrow \mathcal{S}_{post}$} \label{alg:e}
                   \EndIf
                   \State {$\boldsymbol{R}_{d} \leftarrow$ 1}  
               \EndWhile
               \State \Return {$\boldsymbol{R}$}
           \EndFunction
       \end{algorithmic}
   \end{algorithm}
\end{figure}

\subsubsection{Dedicated Per-Replica Ring Buffer}
\label{subsubsec:ring_buf}

\sys relies on the Traffic Orchestrator (\tor) to allocate buffers from the application's memory region comprised by a pool of fixed-size packet buffers (\eg 1500B). 
Concurrent access to the same buffer by multiple stages introduces latency overhead on locking that we must avoid. 
Thus, \sys assigns each stage a set of dedicated per-replica ingress ring buffers.
That is, the number of ingress ring buffers equals to the preceding stage's replication factor in $\boldsymbol{R}$ from Algorithm~\ref{alg:pipe_para}.
Figure~\ref{fig:ring} shows an example where the replication factor of two stages \texttt{S1} and \texttt{S2} are 2 and 3, respectively. 
\tor allocates two dedicated ingress ring buffers for each \texttt{S2} replica to store \texttt{S1}'s egress traffic. 
Then each \texttt{S2} replica processes flows from the two ingress buffers in a round-robin manner.
Moreover, each replica of the last stage has one dedicated egress ring buffer to store the final output.

\subsection{Traffic Orchestration}
\label{subsec:tr}

Now we present the design of traffic redirection and state migration across multiple pipelines.

\begin{figure}[t]
    \centering
    \includegraphics[width=\linewidth]{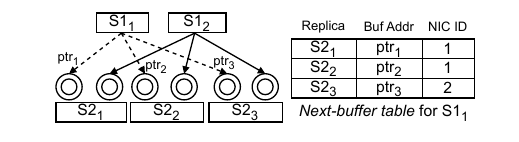}
    \vspace{-10mm}
    \caption{An example of ingress ring buffer assignment {for \texttt{S2}}. \texttt{S1} and \texttt{S2} have two and three replications, respectively.
    The arrows denote traffic direction. In the \textit{next-buffer table}, \texttt{Buf Addr} and \texttt{NIC ID} denotes the buffer address and corresponding \snic location, respectively.}
    \label{fig:ring}
\end{figure}

\noindent{\bf Traffic partitioning and redirection.}
\sys leverages \tor to monitor and redirect per-application traffic.
After resource allocation and application placement (\cref{sec:controlplane}), \sys Controller assigns a globally unique identifier to each stage replica, and {maintains its associated \nic's location, and ingress ring buffer addresses.}
Then each replica informs its \textit{preceding} stage's replicas of the corresponding dedicated ingress buffers they should use through the {\tors}, which is maintained at a \textit{next-buffer table} for each stage replica.
For example, in Figure~\ref{fig:ring}, replica \texttt{S$1_1$}'s next-buffer table has entries for three \texttt{S2} replicas, respectively.

\tor maintains a per-application \textit{flow table} to track the mapping between {a flow and the next ingress buffer it should go to. 
As a new flow arrives at a replica of its first stage, the corresponding \tor selects the next stage replica with the least traffic load from the flow table, inserts a new entry to the table, and directs the packets accordingly (using the next-buffer table).
This ensures a flow is always processed by the same sequence of stage replicas without packet reordering.
Each stage replica directly outputs the packets to the next ingress buffer if it is on the current \nic.
Otherwise, it redirects them to the corresponding \nic hosting the next ingress buffer through \tor.}
For example, in Figure~\ref{fig:ring}, three flows from each \texttt{S1} replica are directed to the three replicas of \texttt{S2}.
Note that the traffic redirection using \tor does not make it the bottleneck as it simply passes packets to the next \nic, as shown in \cref{subsec:micro}. 

\noindent{\bf State migration.}
During adaptive scaling --- adding new stage replicas or shutting down existing ones, it is necessary for stateful applications to migrate states for flows that are moving to another replica. 
We employ a lazy migration approach~\cite{db23,lmnfv}. 
When migrating a flow in replica $s$ to replica $p$, \tor halts $s$'s processing on that flow and stores packets in its ingress buffer for a short time. 
If $s$ and $p$ are on different \nics, the state engine replicates the flow states and transfers them and the cached packets via RDMA to $p$'s ingress buffer; 
otherwise, it directly copies the cached packets to $p$'s ingress buffer.
\tor then updates the flow table by changing the next stage's identifier from $s$ to $p$ for affected flows.

\section{\sys Control Plane}
\label{sec:controlplane}

The control plane's main responsibility is resource allocation, that is to determine for an application 1) the overall amount of each resource it should receive, and 2) the pipeline stage-\nic assignment, i.e. which \nics and per-\nic resource allocation to host the pipelines.
We now present their design in this section.
\Cref{subsec:failover} has more details on the failover mechanism.

\noindent{\bf Overall allocation.}
To determine an application's overall resource allocation, \sys first needs to know its performance targets. 
{\sys allows users to specify maximum throughput targets through \texttt{\sys.set\_thr\_target()}.}
For example, \texttt{\sys.set\_thr\_target(App1,10Gbps)} sets a throughput target of 10Gbps for \texttt{App1}. 
Instantaneous traffic demand exceeding 10Gbps would then cause buffer overflow and packet drops, and the user needs to set a higher throughput target in order for \sys to accommodate this. 
{Due to various factors such as network congestion and resource contention that are difficult to handle, we do not consider latency targets in this work.}

With the performance target, our approach is to first understand the application's baseline performance with minimal resources, then allocate enough resources to meet the target. Note when scaling the allocation, we should use the most efficient partial replication strategy for the pipelines as mentioned in \cref{subsub:pipe_para}.

\noindent{\underline{\it Application profiling.}}
We first profile single-pipeline baseline performance by running each CPU-based stage with just one \textit{resource unit} (one core and 4GB memory) and the corresponding accelerators specified by accelerator function APIs in the program. 
This is the \textit{minimal allocation} $\boldsymbol{I}$ for applications.
We measure the baseline end-to-end throughput $t_p$ and latency $l_p$, and the per-stage throughput and latency vectors $\boldsymbol{t_s}$ and $\boldsymbol{l_s}$, respectively. 
Then we employ Algorithm~\ref{alg:pipe_para} with $\boldsymbol{l_s}$ as input to derive the best partial replication strategy $\boldsymbol{R}$ (\cref{subsub:pipe_para}). 
We apply $\boldsymbol{R}$ to the minimal allocation to obtain the \textit{$\boldsymbol{R}$-allocation}, and profile the the corresponding end-to-end throughput $t_R$ and latency $l_R$.

\noindent{\underline{\it Meeting throughput target.}}
Let us consider the common case of throughput target $t_t$. 
If $t_t \leq t_p$, the minimal allocation $\boldsymbol{I}$ is enough (one {resource unit} per CPU stage with specific accelerators). 
If $t_p < t_t \leq t_R$, one {$\boldsymbol{R}$-allocation} is enough. 
When $t_t > t_R$, 
\sys increases allocation in the granularity of $\boldsymbol{R}$-allocation, computing the total per-stage allocation $\boldsymbol{r_s}$ as $\boldsymbol{R}$ via {$\lceil t_t / t_R \rceil\boldsymbol{R}$}.

\begin{figure}[t]
    \vspace{-3mm}
    \small
    \begin{algorithm}[H]
        \centering
        \caption{Locality-aware pipeline stage-\nic assignment}
        \label{alg:lara}
        \begin{algorithmic}[1]
            \State {$\mathcal{S}$: List of pipeline stages; $\mathcal{N}$: List of \snics}
            \State {$\boldsymbol{r_s}$: Per-stage total resource demands}
            \State {$\boldsymbol{r_{nic}}$: Available resources of \snics}
            \State {$\boldsymbol{A}$: Current resource allocation results}
            \Function {RESOURCE\_ALLOC}{$\mathcal{S}$, $\mathcal{N}$, $\boldsymbol{r_s}$, $\boldsymbol{r_{nic}}$, $\boldsymbol{A}$}
                \For {$s$ in $\mathcal{S}$} 
                    \State {$s^+ \leftarrow $ the preceding stage of $s$} \label{galg:a1}
                    \While {$\boldsymbol{r_s}[s] > 0$}
                        \State {$n \leftarrow $ FIND\_NEXT\_NIC($\mathcal{N}$, $\boldsymbol{r_{nic}}$, $s^+$)} \label{galg:c1}
                        \State {$\boldsymbol{A} \leftarrow$ ALLOC\_ONE\_NIC($\boldsymbol{r_s}$, $\boldsymbol{r_{nic}}$, $\boldsymbol{A}$, $n$, $s$)}
                    \EndWhile
                \EndFor
                \State \Return {$\boldsymbol{A}$}
            \EndFunction
            \Function {FIND\_NEXT\_NIC}{$\mathcal{N}$, $\boldsymbol{r_{nic}}$, $s^+$}
            \State {$\mathcal{N} \leftarrow$ sort($\mathcal{N}$, keys=(distance($s^+$),  $\boldsymbol{r_{nic}}$))}\label{galg:d1} 
                \For {$n$ in $\mathcal{N}$} \label{galg:d2}
                    \If {$\boldsymbol{r_{nic}}[n] \leq 0$} {\footnotesize /* No available resources */}
                        \State {\textbf{continue}}
                    \Else
                        \State {\Return {$n$}} \label{galg:d3}
                    \EndIf
                \EndFor
            \EndFunction
            \Function {ALLOC\_ONE\_NIC}{$\boldsymbol{r_s}$, $\boldsymbol{r_{nic}}$, $\boldsymbol{A}$, $n$, $s$}
                \If {$\boldsymbol{r_s}[s] \geq \boldsymbol{r_{nic}}[n,\gamma_s]$} \label{galg:hhh7}
                    \State {$\boldsymbol{A}[n,s] \mathrel{+}=\boldsymbol{r_{nic}}[n,\gamma_s]$, $\boldsymbol{r_s}[s] \mathrel{-}= \boldsymbol{r_{nic}}[n,\gamma_s]$}
                    \State {$\boldsymbol{r_{nic}}[n,\gamma_s]=0$}\label{galg:hhh5}
                \Else
                    \State {$\boldsymbol{A}[n,s] \mathrel{+}=\boldsymbol{r_s}[s]$, $\boldsymbol{r_{nic}}[n,\gamma_s]\mathrel{-}=\boldsymbol{r_s}[s]$}\label{galg:hhh6}
                    \State {$\boldsymbol{r_s}[s]=0$}
                \EndIf
            \EndFunction
            
        \end{algorithmic}
    \end{algorithm}
\end{figure}

\noindent{\bf Pipeline stage-\nic assignment.}
\sys then needs to decide which \nic to host each pipeline stage given $\boldsymbol{r_s}$.
This is challenging due to a few additional factors we must consider. 
Distributing pipeline stages across \nics inevitably introduces network overhead in both bandwidth and latency which need to be minimized.
Also, \snics have heterogeneous available resources due to different pipelines they already host before this new application. 

We rely on a \textit{locality-aware} heuristic to take into account these factors. 
Our approach consolidates consecutive stages onto as few \nics as possible to reduce traffic redirection overhead in the network.
\sys Controller maintains an assignment matrix $\boldsymbol{A}$ {for each application} to track the current assignment, where $\boldsymbol{A}[n,s]$ denotes the allocated resources of stage $s$ on \nic $n$.
These resources can be \textit{resource units} or accelerators depending on the stage.

Algorithm~\ref{alg:lara} outlines our assignment solution.
With per-stage ($\boldsymbol{r_s}$ and $t_s$) and per-\nic information ($\boldsymbol{r_{nic}}$), 
\sys finds best \nics to host each stage iteratively. 
For each stage $s$, we create the candidate \nic list (line~\ref{galg:c1}) by sorting \nics based on its distance to the \nic that hosts $s$'s preceding stage $s^+$, and the \nic's available resources (line~\ref{galg:d1}).
Then the first candidate \nic $n$ with available resources (lines~\ref{galg:d2}-\ref{galg:d3}) is chosen (for this iteration). 
Next, \texttt{ALLOC\_ONE\_NIC()} assigns resources from $n$ to $s$.
If $n$'s available resources $\boldsymbol{r_{nic}}[n,\gamma_s]$ can meet its demand $\boldsymbol{r_s}[s]$ on stage $s$'s resource $\gamma_s$ (line~\ref{galg:hhh7}), allocation is completed.
Otherwise, we can only allocate $\boldsymbol{r_{nic}}[n,\gamma_s]$ to $s$.
Finally, we update the assignment $\boldsymbol{A}$ and $s$'s demand $\boldsymbol{r_s}[s]$ accordingly.

\noindent{\textbf{Adaptive scaling.}}
When a new throughput target is set, \sys calculates the new per-stage resource demands $\boldsymbol{\hat{r}_s}$ with Algorithm~\ref{alg:pipe_para} and adjusts the allocation accordingly. 
{In case $\boldsymbol{\hat{r}_s} > \boldsymbol{r_s}$}, \sys allocates additional resources by Algorithm~\ref{alg:lara} while maintaining the current pipelines, and updates up the flow tables through \tor. 
If $\boldsymbol{\hat{r}_s} < \boldsymbol{r_s}$, \sys de-allocates existing resources by adjusting \tor's redirection policies, migrating flows to the remaining pipelines, and finally reclaiming the associated resources with the unnecessary stages.

\section{Implementation}
\label{sec:implementation}

\begin{table}[t]
    \centering
    \resizebox{\columnwidth}{!}{
    \begin{tabular}{lccc}
    \toprule
    App. & Abs. & \# & Resources \\ 
    \midrule
    Intrusion Detection (ID) \cite{e319,ipsdoca} & packet   & 3 & CPU, regex \\
    IPComp Gateway (ICG) \cite{ipcomprfc,nfp17} & packet   & 2 & CPU, compression \\
    IPsec Gateway (ISG) \cite{e319,dyk23} & packet  & 4 & CPU, regex, AES \\  
    Firewall (FW) \cite{wl1,dyk23} & packet  & 2 & CPU \\ 
    Flow Monitor (FM) \cite{e319,wl1} & packet & 2 & CPU \\
    L7 Load Balancer (LLB) \cite{nfp17,microboxes18} & socket & 1 & CPU \\ 
    \bottomrule
    \end{tabular}
    }
    \caption{Applications in our evaluation. ICG and ISG are stateless, while others are stateful. \# is the number of {\uco}s for the app.} 
    \label{table:wl}
    \vspace{-5mm}
\end{table}

We prototype \sys on NVIDIA BlueField-1 \cite{mbf1}, BlueField-2 \cite{mbf2}, and AMD Pensando \snics \cite{pensando}.  
It currently does not utilize the P4 ASIC on Pensando.
The implementation is based on the default C SDK of these \nics, uses $\sim$4000 LoC on Ubuntu 20.04 with Linux kernel 5.4.0, and exposes a dynamic shared library \texttt{lib\sys}. 
Each executor runs as a Docker container, and can generally launch multiple (sub-)pipelines. 
\texttt{cgroup} and \texttt{namespace} are used for resource allocation. 
Communication among \ca, \tor, and executors on the same \nic uses shared memory \cite{zhixiong20netkernel, bojie2020socksdirect}. 
\sys relies on DPDK and F-stack \cite{f-stack} for packet and socket processing, respectively, and the \tor uses DPDK for traffic redirection. 
State engine uses a linked hash table with 4096 buckets.  
For multiple applications, \sys follows a \textit{first-come-first-serve} way, 
and reclaims an application's resources upon its termination.

\noindent{\bf Monitoring.}
In the per-\nic CA, Resource Manager and Runtime Manager operate as separate threads.
The Resource Manager tracks local resource availability using standard monitoring tools like \texttt{Perf} and \texttt{htop}, while the Runtime Manager monitors pipeline execution statistics (\eg latency) by tracing and logging requests or traffic through \tor, which is done periodically and outside the data path. For example, \tor captures request latency by noting timestamps at reception and transmission. \sys Controller communicates with {\ca}s via RDMA channels.

\noindent{\bf Applications.}
We implement six popular \snic applications on \sys, as shown in Table~\ref{table:wl}, with $\sim$3900 LoC in total \cite{opensource}.  
They incorporate various \ucfs running on ARM cores and hardware accelerators.
Among them, L7 Load Balancer employs \sys's socket processing operations, while the remaining uses \sys's packet processing operations.
For stateful applications, Flow Monitor utilizes the \texttt{COMPUTE} operator to calculate aggregated flow metrics (\eg packet counts). 
The other three use \texttt{ADD}, \texttt{REMOVE}, \texttt{GET}, and \texttt{SET} operators.

\section{Evaluation}
\label{sec:evaluation}

\begin{figure*}[t]
    \begin{minipage}[t]{0.33\linewidth}
        \vspace{0pt}
        \centering
    \includegraphics[width=\linewidth]{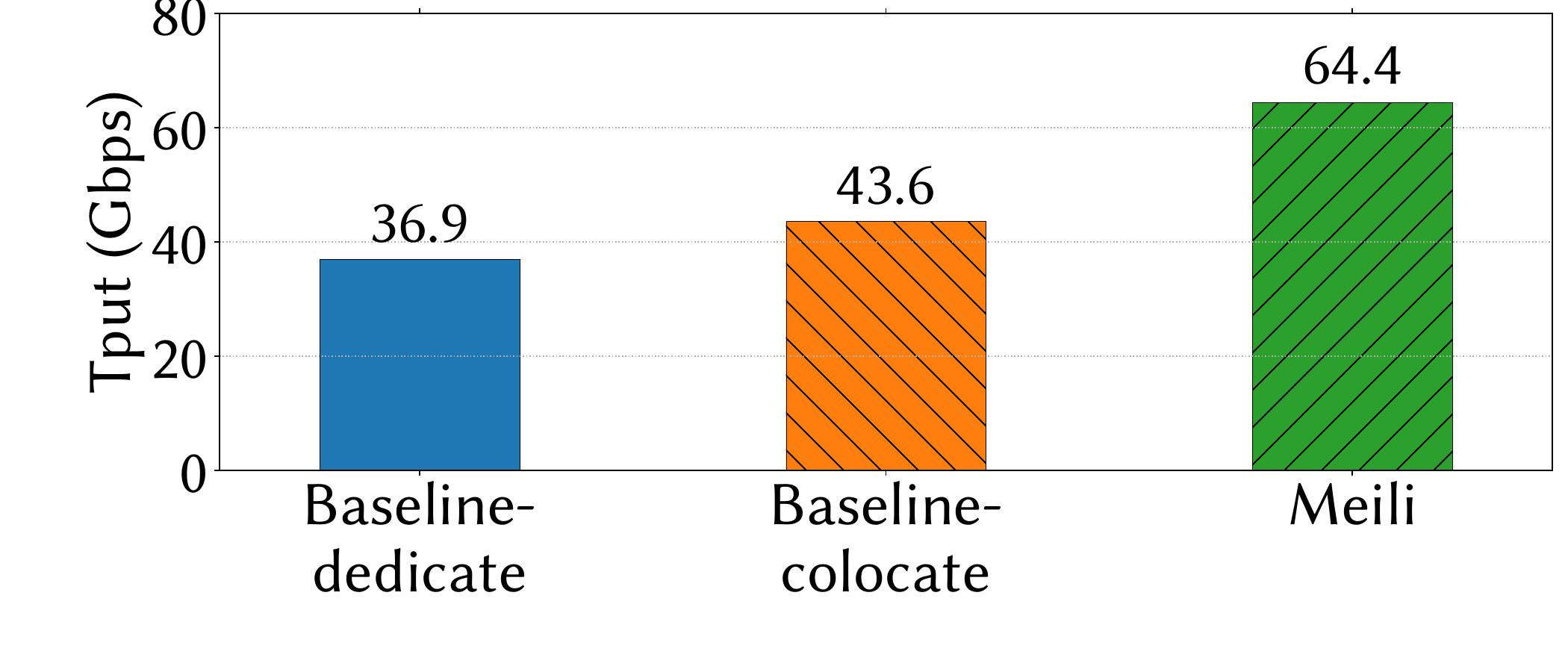}
        \vspace{-10mm}
        \caption{Maximum achievable per-app throughput comparison in Cluster 1.}
        \label{fig:result_reff}
    \end{minipage}
    \hfill
    \begin{minipage}[t]{0.33\linewidth}
        \vspace{0pt}
        \centering
        \includegraphics[width=\linewidth]{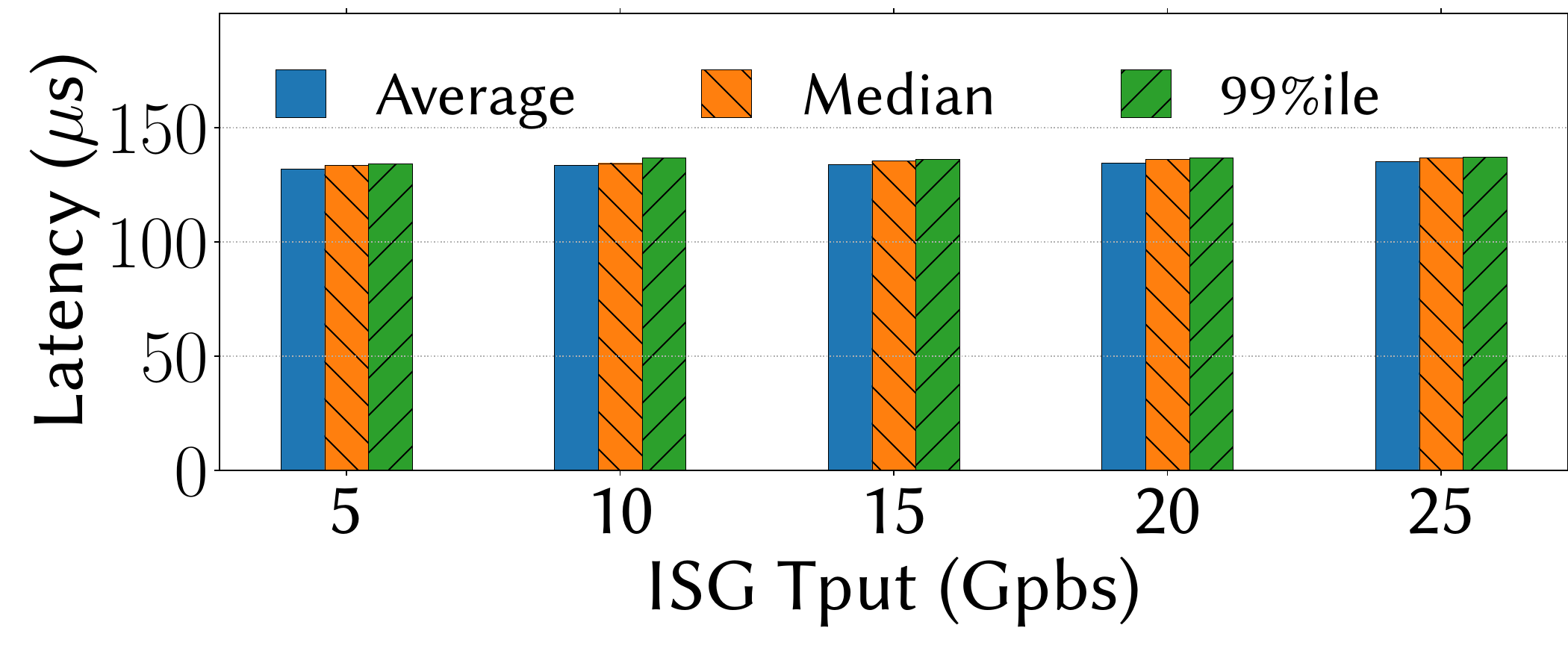}
        \vspace{-10mm}
        \caption{ISG latency with different throughput targets in Cluster 2.}
        \label{fig:corun_lat}
    \end{minipage}
    \hfill
    \begin{minipage}[t]{0.33\linewidth}
        \vspace{0pt}
        \centering
        \includegraphics[width=\linewidth]{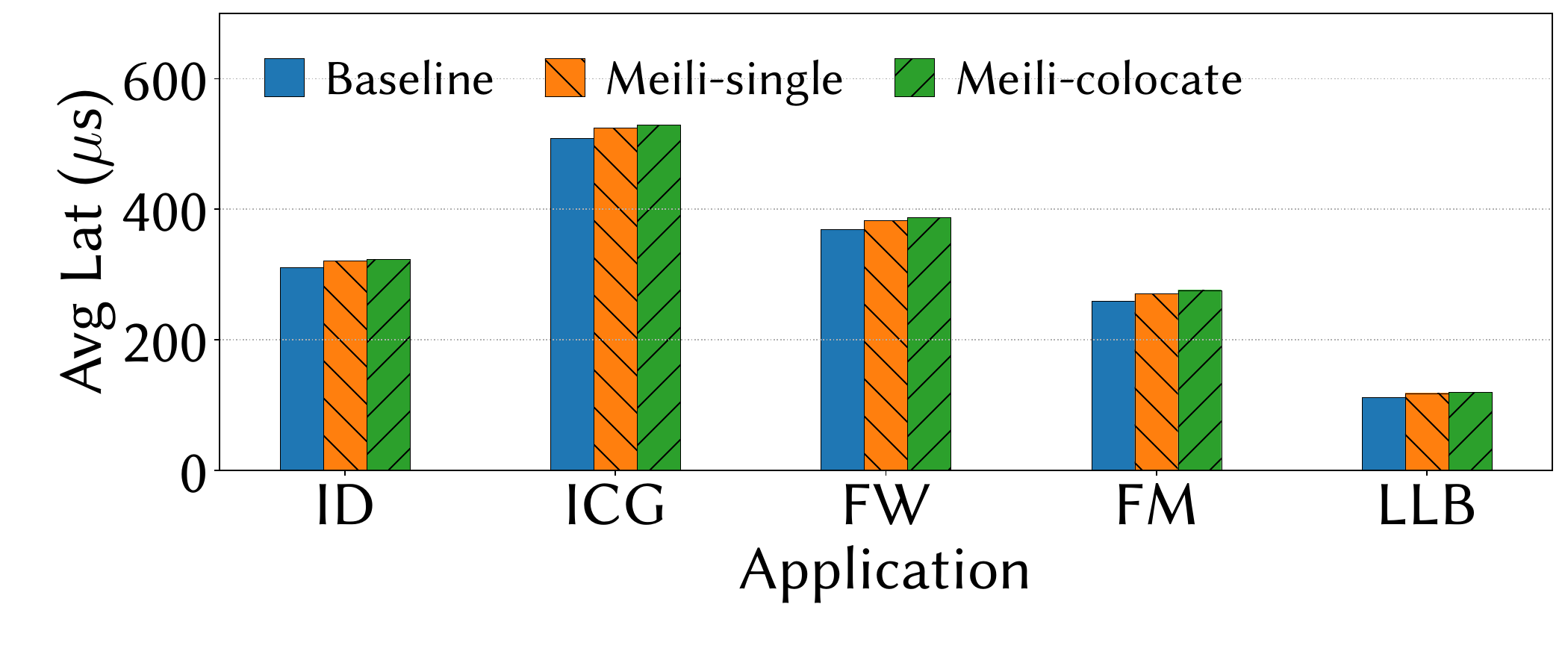}
        \vspace{-10mm}
        \caption{App latency when co-running in Cluster 1.}
        \label{fig:result_reff_lat}
    \end{minipage}
\end{figure*}

Our evaluation highlights \sys's benefits in: 
(a) attaining up to 1.75$\times$ better resource efficiency than current deployment approaches \cite{dyk23,db23,e319,le17uno} (\cref{subsec:reseff});
(b) enhancing resource availability (\cref{subsec:resava});
(c) improving host resource saving in a new usecase for 5G UPF offloading (\cref{subsec:5gupf}). 
The results also confirm that our design is effective in:
(a) achieving scalable app throughput with very low latency overhead  (\cref{subsec:micro});
and (b) supporting adaptive scaling (\cref{subsec:dynare}).

\noindent{\bf Methodology.}
We build a cluster with 8 servers in the same rack. 
Each server has two 100GbE \snics connected to a Mellanox SN2700 switch \cite{snswitch}.
Our cluster uses 8 BlueField-2 (\bftwo) \cite{mbf2} (each with 8 ARMv8 A72 cores, 1 regex and 1 compression accelerator), 
4 BlueField-1 (\bfone) \cite{mbf1} (each with 16 ARMv8 A72 cores), 
and 4 Pensando \snics \cite{pensando} (each with 16 ARMv8 A72 cores, 1 AES and 1 compression accelerator). 
All onboard ARMv8 A72 cores exhibit uniform per-core performance. 
Each client has an AMD EPYC-7542 CPU with 32 cores at 2.9GHz, 256GB DRAM, and a ConnectX-6 100GbE NIC.
Latency results are obtained over 5000 runs after enough warmup rounds. 
We use the six common applications (apps) listed in Table~\ref{table:wl}, and DPDK-Pktgen to generate flows with 1500B packets.

\subsection{Benefit 1: Resource Efficiency}
\label{subsec:reseff}

\sys improves cluster-wise resource efficiency through resource pooling and fine-grained orchestration. 
We show this benefit by deploying all apps
in a cluster with 4 \bfone and 8 \bftwo (Cluster 1) except ISG (AES accelerator is only on Pensando).
We generate 128 flows for each app, with the flow size following the uniform distribution, and identical flows are used in each run for all systems.
We compare \sys to two typical resource scaling approaches. 
Both deploy and scale the entire app in a \textit{per-instance} granularity, with each instance running entirely on one \nic, but differ in how they share a \nic: 
(1) \textit{Baseline-dedicate}, the default approach in most existing systems like Sirius \cite{db23}, which dedicates each \nic to one app instance \cite{dyk23,db23,e319};
(2) \textit{Baseline-colocate}, which allows multiple instances to share a \nic \cite{le17uno,panic20,supernic22,nica19,fairnic20}. 
To compare resource efficiency, we set a uniform throughput target for each app (for fairness) starting from the \nic line rate, and investigate the maximum throughput each system can satisfy.
If a target cannot be met (even for just one app), we lower the target by 100Mbps for all apps and retry.

Figure~\ref{fig:result_reff} shows the maximum achievable throughput results.
\sys achieves 69.4Gbps per app, 1.75$\times$ and 1.48$\times$ that of \textit{Baseline-dedicate} and \textit{Baseline-colocate}, respectively. 
\textit{Baseline-colocate} benefits from sharing the \nic by multiple apps. 
For example, \textit{Baseline-dedicate} allocates 3 \bftwo to ID, achieving 36.9Gbps. 
In \textit{Baseline-colocate}, \bftwo ID utilizes 3 extra resource units on another \bftwo which is shared with ICG in addition to the 3 \bftwo and attains 43.6Gbps, while this \nic is dedicated to ID in \textit{Baseline-dedicate}. 
\sys enjoys similar benefits in this aspect.
Moreover, to see \sys's benefit over \textit{Baseline-colocate}, 
note that in \textit{Baseline-colocate} ID and ICG can only be deployed to \bftwo (Table~\ref{table:wl}) {since \bfone does not have any accelerators}, while in \sys they can also utilize \bfone's CPUs to further improve utilization due to its \textit{per-pipeline stage} scaling (\cref{subsub:pipe_para}).  
The latency overheads of distributed deployment is minimal as we will show in \cref{subsubsec:app_perf}.

\subsection{Benefit 2: Resource Availability}
\label{subsec:resava}

In this experiment we launch IPsec Gateway (ISG) that uses regex accelerator on \bftwo and AES accelerator on Pensando simultaneously, and scale its resources to achieve throughput targets from 5Gbps to 25Gbps, in a cluster with 4 \bfone, 8 \bftwo, and 4 Pensando (Cluster 2). 
ISG's pipeline is partitioned into two parts: one sub-pipeline utilizing CPU and regex accelerator on \bftwo, and another running on CPU and AES accelerator on Pensando. 
The traffic setting is the same as that in \cref{subsec:reseff}.
Figure~\ref{fig:corun_lat} shows ISG's latency statistics in \sys. 
Observe that the latencies remain stable as the resources scale with the throughput target, proving its scalability in \sys. Note that \tor adds an average latency overhead of $\sim$9.2\us. 
This shows the improved availability brought by \sys compared to \textit{Baseline-colocate} and \textit{Baseline-dedicate}, which cannot deploy ISG in this case with the per-pipeline granularity as no single \nic in Cluster 2 has all the required resources.

\subsection{Benefit 3: Resource and Cost Savings}
\label{subsec:5gupf}

We show this benefit with a new usecase which involves the User Plane Functions (UPFs) in cloud-based 5G core networks \cite{ericlo,hwran,nvdran,l25gc}. Offloading UPFs to programmable hardware can free up host CPU cores for tenants \cite{xplan,accupf}. 
\sys enhances this by efficiently multiplexing heterogeneous onboard resources. We implement three typical UPFs \cite{5gov} --- Traffic Usage Reporting (TUR), DDoS, and Packet Inspection (PI) --- on a \bftwo and a Pensando \snic. 
All UPFs need onboard CPU and PI also utilizes regex and AES accelerators. 
We compare \sys against two typical deployment approaches for achieving 20Gbps per UPF:
(1) \textit{Host-only}, relying solely on host CPU; (2) \textit{Baseline-offload}, which offloads UPFs on individual \nics. 
The traffic comprises of 16 flows with uniformly-random flow sizes. 
In \textit{Host-only}, TUR, DDoS, and PI demand 4, 3, and 8 host CPU cores, respectively, with regex and AES implemented by software approaches HyperScan \cite{xw19} and OpenSSL \cite{openssl}. 
\textit{Baseline-offload} saves 7 host CPU cores by offloading TUR and DDoS with 6 and 5 onboard CPU cores on \bftwo and Pensando, respectively. 
Yet PI cannot be offloaded due to lack of both accelerators on either \nic here. 
\sys keeps the same deployment for TUR and DDoS as \textit{Baseline-offload}, and enables PI offloading through two sub-pipelines: one using regex accelerator on \bftwo and the other utilizing AES accelerator and 5 onboard CPU cores on Pensando. 
This way \sys saves 8 more host cores compared to \textit{Baseline-offload}, with each core generating a potential revenue of around USD \$900/yr \cite{FPMC18}.

\begin{figure}[t]
    \centering
    \includegraphics[width=\linewidth]{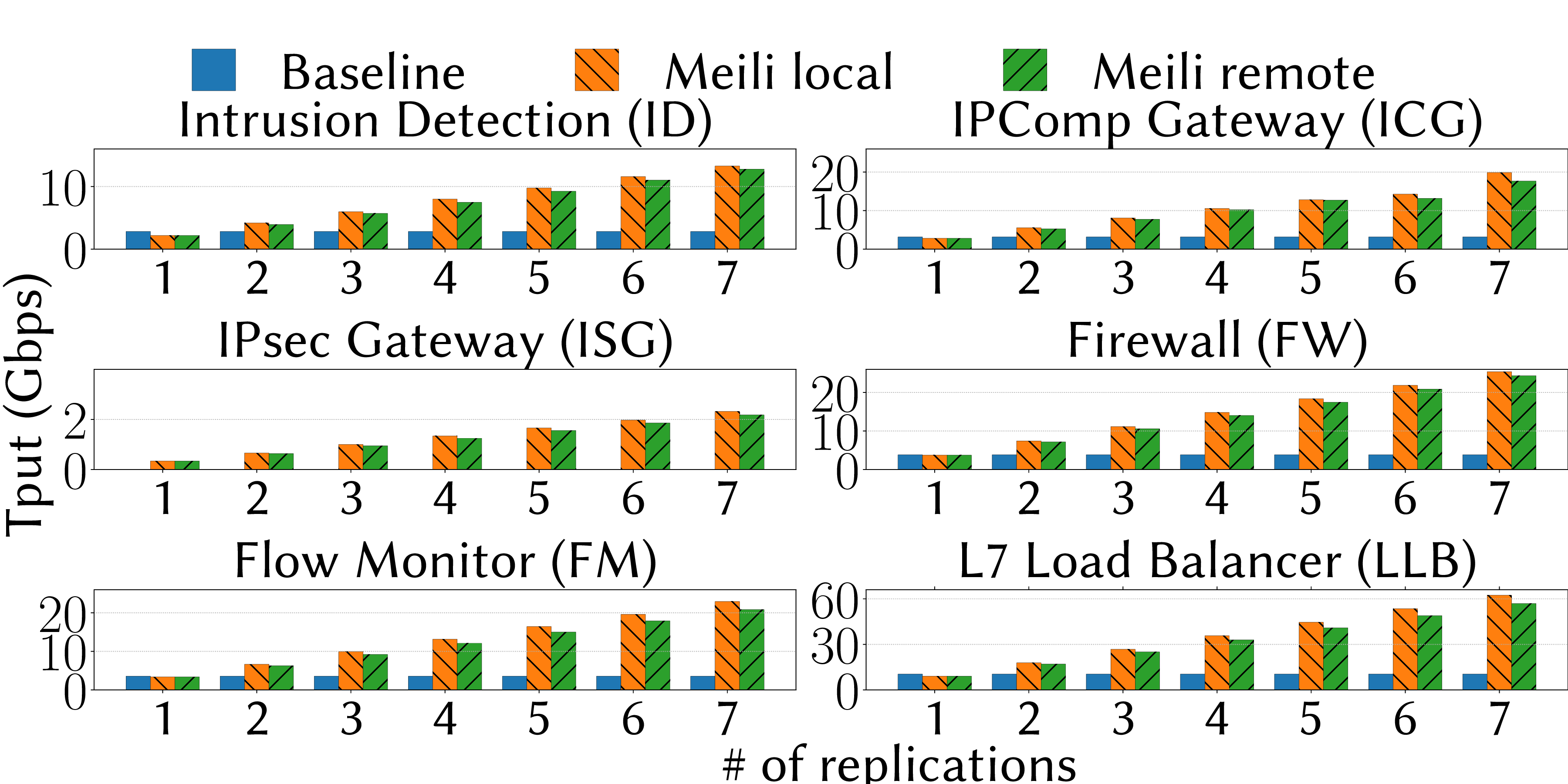}
    \vspace{-5mm}
    \caption{App end-to-end throughput of scaling pipelines.
    ``\sys local'' scaling pipelines only in the local \nic, while ``\sys remote'' scales a new pipeline on a new \nic each time. 
    One \bftwo can host at most 7 pipelines (with one resource unit each). 
    For ISG, the sub-pipeline on \bftwo is scaled both locally and remotely, while the sub-pipeline on Pensando remains in the same \nic.
    }
    \label{fig:result_fw_scale}
\end{figure}

\subsection{Microbenchmarks}
\label{subsec:micro}

We now look at \sys's microbenchmarks in terms of performance and overhead.

\begin{figure*}[t]
    \begin{minipage}[t]{0.33\linewidth}
        \vspace{0pt}
        \centering
        \includegraphics[width=\linewidth]{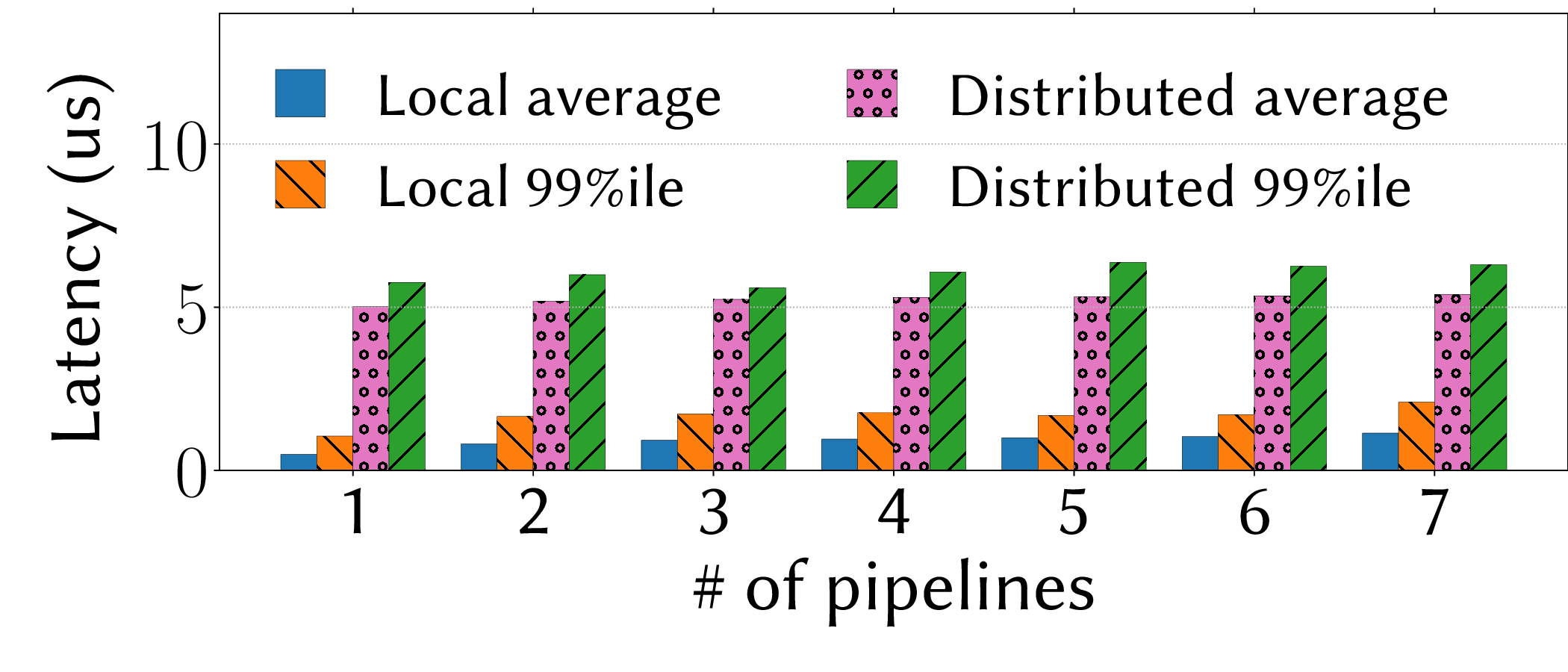}
        \vspace{-10mm}
        \caption{End-to-end latencies of traffic partitioning.}
        \label{fig:re_pl}
    \end{minipage}
    \hfill
    \begin{minipage}[t]{0.33\linewidth}
        \vspace{0pt}
        \centering
        \includegraphics[width=\linewidth]{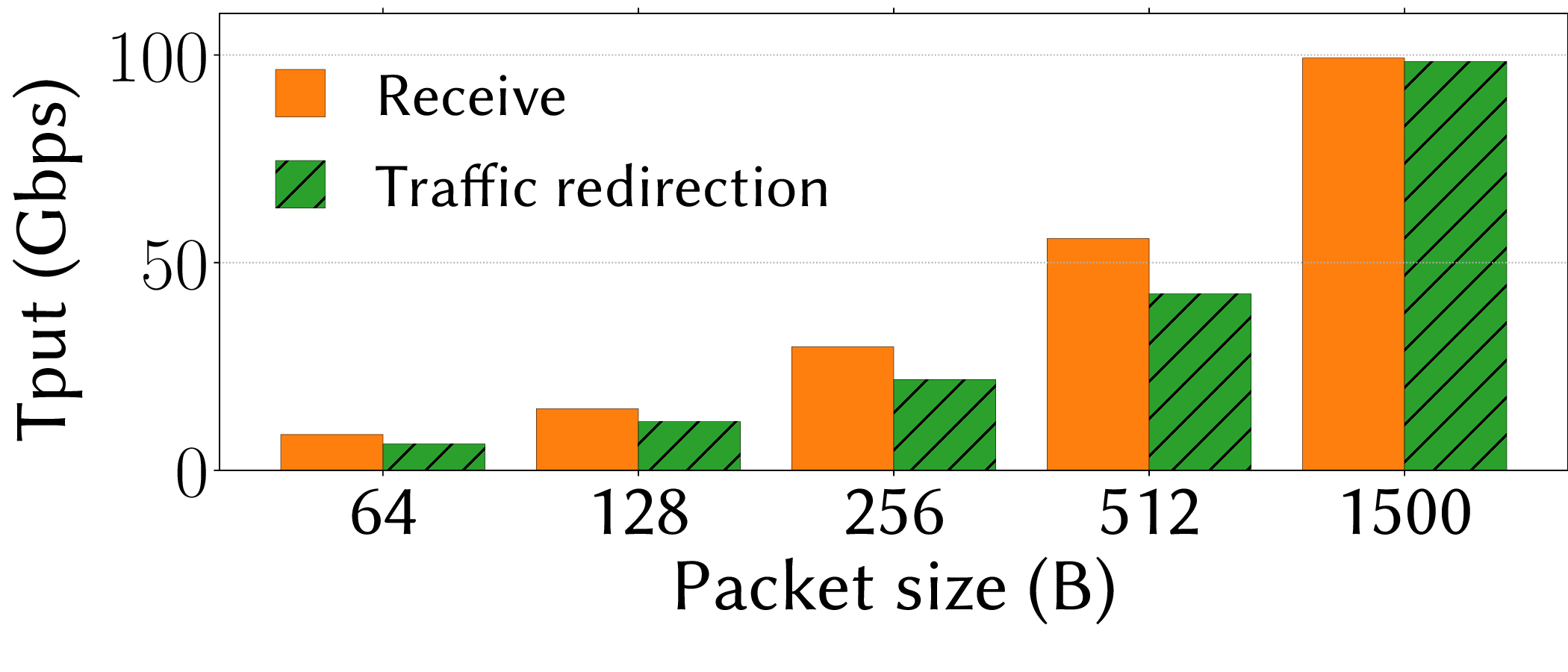}
        \vspace{-10mm}
        \caption{Single-core traffic redirection throughput when the packet size increases.}
        \label{fig:result_to_raw_thr}
    \end{minipage}
    \hfill
    \begin{minipage}[t]{0.33\linewidth}
        \vspace{0pt}
        \centering
        \includegraphics[width=\linewidth]{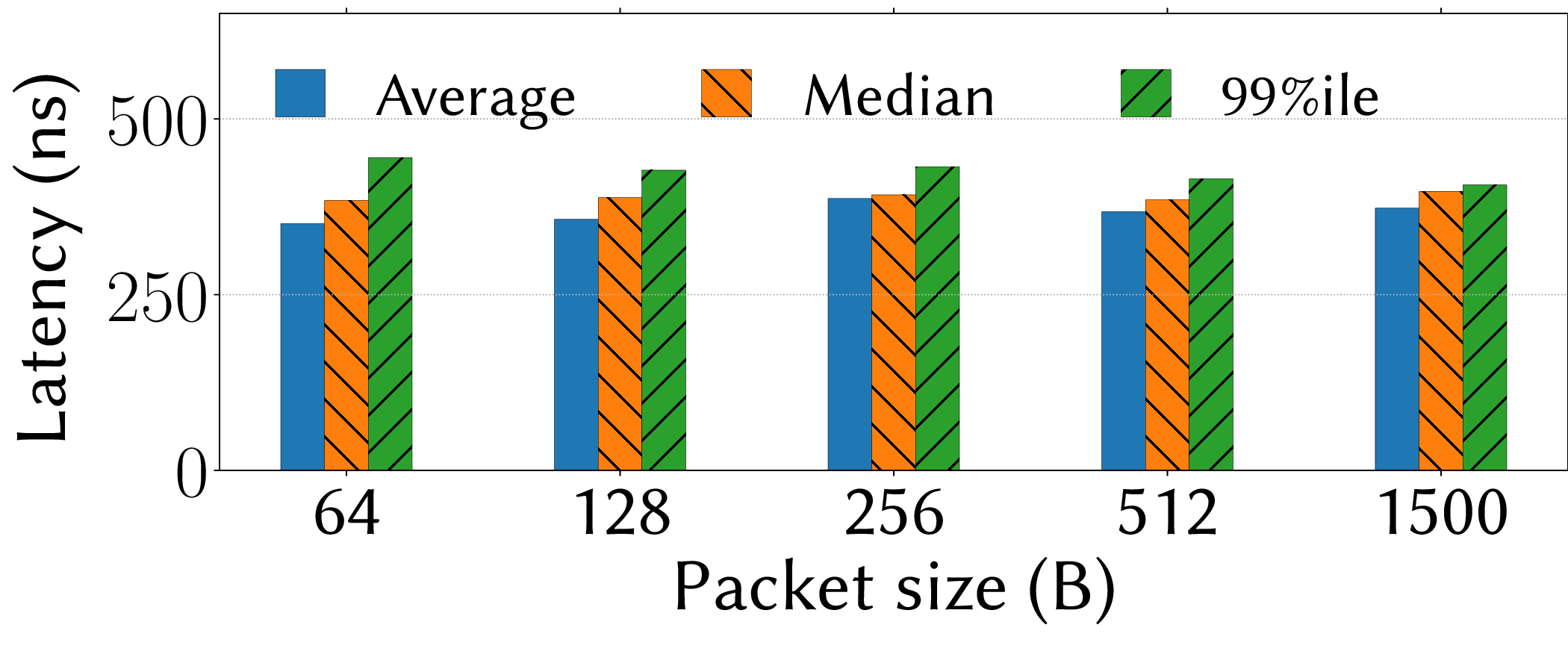}
        \vspace{-10mm}
        \caption{Per-packet latencies of traffic redirection.}
        \label{fig:to_lat}
    \end{minipage}
\end{figure*}

\subsubsection{App Performance}
\label{subsubsec:app_perf}

We first show end-to-end latency and throughput scalability of apps.

\noindent{\bf End-to-end latency.} 
We measure the average latencies when deploying apps in Cluster 1.
All settings are the same as in \cref{subsec:reseff}.
Figure~\ref{fig:result_reff_lat} shows the results.
Here \textit{Baseline} denotes the case where each app runs independently on a single \nic. 
In \textit{\sys-single} and \textit{\sys-colocate}, each app is deployed in Cluster 1 using the same allocation as in Figure~\ref{fig:result_reff}; 
Apps run independently and concurrently in \textit{\sys-single} and \textit{\sys-colocate}, respectively.
We see that \textit{\sys-colocate} incurs minor latency increases over \textit{\sys-single}, mainly due to \sys's resource partitioning for multiple apps. 
This illustrates \sys's tradeoff between resource efficiency from colocation and app latency interference. 
We expect emerging resource isolation approaches~\cite{fairnic20, KRFX18} can effectively address this issue.
In addition, \textit{\sys-single} introduces a slight latency increase compared to \textit{Baseline}; 
when distributing app pipeline stages in the cluster, another $\sim$6\us to $\sim$15\us increase is observed, which is 2.9\% to 5.4\% of the end-to-end app latency. 
This is attributed to the overhead on remote state access for stateful apps and inter-stage traffic redirection by \tor (\cref{subsubsec:traf_orch}). 
Compared to the local state access in \textit{Baseline}, the remote state access of ID, FW, FM, and LLB in \textit{\sys-single} incurs an additional 5.2\us, 2.9\us, 3.3\us, and 4.1\us, respectively.

\noindent{\bf Throughput scalability.}
\sys's data-path parallelism allows apps to attain scalable throughput by replicating multiple pipelines and distributing flows across them (\cref{subsec:dpp}).
To highlight this, we measure the end-to-end app throughput of scaling pipelines on a single \nic and across multiple \nics. 
We allocate one resource unit for all CPU-based stages together within every pipeline replica.
Accelerators here are not throughput bottleneck in our experiment so we only scale CPU-based stages.
We compare \sys against conventional standalone deployment (Baseline), 
which only processes flows in a single pipeline on one \nic. The traffic setting is the same as that in \cref{subsec:reseff}.
Each \nic uses one CPU core for \tor. 
Figure~\ref{fig:result_fw_scale} shows the results. 
We can see Baseline clearly lacks scalability while \sys attains scalable throughput by concurrently processing flows across multiple pipelines. 
The gain is significant: For example with 7 pipelines, Firewall and Flow Monitor achieve $\sim$25Gbps, and L7 Load Balancer achieves $\sim$60Gbps.
On one pipeline, \sys's throughput is slightly lower than Baseline due to its traffic partitioning overhead. 
When replicating pipelines over multiple \nics, \sys has a throughput drop of $\sim$5\% to $\sim$10\% compared to doing that locally, mainly due to the use of multiple \tor. 
Note that ISG cannot be deployed in Baseline as no single \nic has all the required resources, but \sys enables it by leveraging CPU cores and accelerators on \bftwo and Pensando simultaneously.

\begin{table}[t]
        \centering
        \resizebox{0.9\columnwidth}{!}{
        \begin{tabular}{lcccccc}

        \toprule
        \# of states & $2^6$ & $2^8$ & $2^{10}$ & $2^{12}$ & $2^{14}$ & $2^{16}$ \\ 
        \midrule
        Local READ & 31 & 42 & 162 & 735 & 4953 & 80466 \\
        Local WRITE & 107 & 124 & 255 & 836 & 3149 & 12536 \\ 
        Remote READ & 34 & 52 & 189 & 830 & 5319 & 81919 \\ 
        \bottomrule
        \end{tabular}
        }
        \caption{The latencies (\us) of reading and writing states.}
        \label{table:state_lat}
\end{table}

\subsubsection{Traffic Orchestrator}
\label{subsubsec:traf_orch}

We present the overhead on \tor, including traffic partitioning and redirection.

\noindent{\bf Traffic partitioning.}
We evaluate end-to-end latencies of partitioning flows across pipelines,  
each with only a single stage on a resource unit, transferring packet buffer pointers from ingress to egress ring buffers. 
The traffic includes 128 flows with flow size following uniform distribution, and each \tor runs with one core. 
Figure~\ref{fig:re_pl} shows latencies for two cases: 
(1) pipelines on the same \nic, and (2) pipelines across \nics, each hosting only one pipeline. 
Observe the latencies remain at several \us, proving low overhead in traffic partitioning. 
However, the distributed case has a higher latency of $\sim$4.52\us due to extra inter-\nic round-trips.

\noindent{\bf Traffic redirection.}
We first examine \tor's traffic redirection performance by measuring single-core throughput with varying packet sizes.
Figure~\ref{fig:result_to_raw_thr} shows the results, revealing that \tor achieves line-rate redirection (100Gbps) at 1500B. 
However, a roughly 20\% reduction occurs for small packet sizes due to per-packet processing overhead. 
We also examine the per-packet processing latency of traffic redirection as in Figure~\ref{fig:to_lat}. 
The results indicate that redirection introduces sub-$\mu$s overhead, primarily attributed to attaching unique sequences to each packet and rerouting. 
Additionally, average latency remains stable with increasing packet sizes, thanks to \tor's forwarding based on packet headers (5-tuple).

\subsubsection{State Operation}
We examine state operation latencies in \sys. Each state has a unique random \texttt{h\_key}. 
Table~\ref{table:state_lat} presents average end-to-end latencies for local state read, and local and remote state write, fundamental for \texttt{GET}, \texttt{SET}, \texttt{ADD}, and \texttt{REMOVE} operations. 
As state count grows, reading latencies exceed writing due to the \texttt{h\_key} comparisons during hash collisions of reading. 
This overhead can be mitigated by using more buckets. 
We also measure \texttt{TRAVERSE} and \texttt{COMPUTE} latencies across 8 \snics, each housing $2^{16}$ states. The operator of \texttt{COMPUTE} is addition.
We see that \texttt{TRAVERSE} and \texttt{COMPUTE} take 10.69\ms and 64\us, respectively.
This gap is because \texttt{TRAVERSE} requires reading all states from remote \snics, while \texttt{COMPUTE} only requires the transmission of addition instructions and aggregated results.

\begin{figure}[t]
    \centering
    \includegraphics[width=0.85\linewidth]{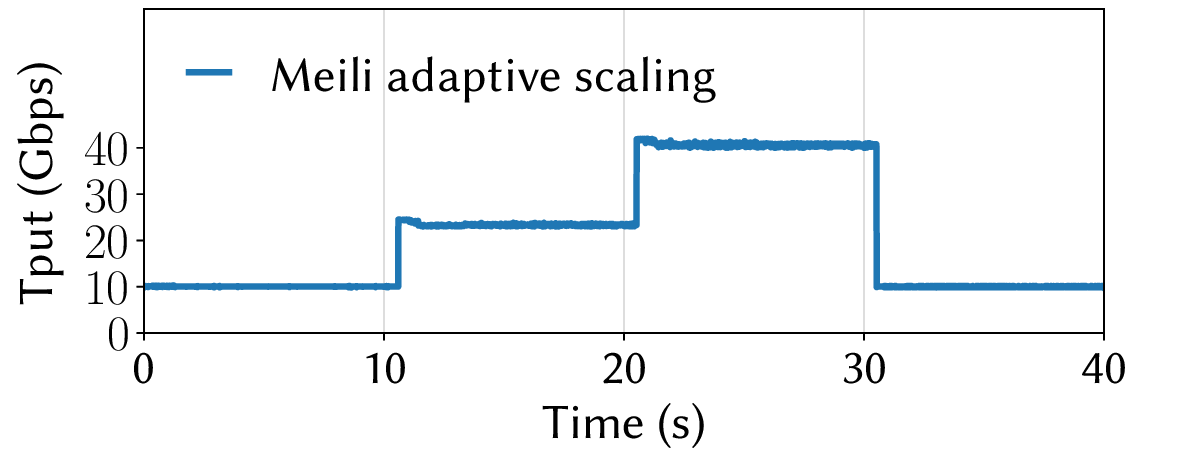}
    \vspace{-3mm}
    \caption{Firewall under dynamic throughput targets.}
    \label{fig:dynamic_req}
\end{figure}

\subsection{Adaptive Scaling}
\label{subsec:dynare}

We examine \sys's adaptive scaling now.
Initially, we run a Firewall with 32 flows and a 10Gbps target in our testbed.
During runtime, we adjust the target to 20Gbps and 40Gbps, and then back to 10Gbps at 10s, 20s, and 30s. 
Figure~\ref{fig:dynamic_req} shows that \sys efficiently meets dynamic throughput targets with $\sim$400ms response time.

\section{Discussion}
\label{sec:discussion}

This section discusses some immediate concerns about \sys.

\noindent{\bf Is \sys applicable to all \snics?}
In our vision, an ideal \sys implementation should support all \snic architectures, including SoC, FPGA, P4 ASICs, etc. 
Achieving this requires developing a compiler that compiles applications down to all these architectures, which is time- and resource-consuming and beyond the scope of this work.  
As a result, \sys's current design focuses on SoC-based \snics, and the programming model aligns with the SoC architecture. 
In case FPGA or P4 ASICs are used to run fixed functions, they can be managed as additional accelerators in \sys's pool. 
For SoC \snics with different processor architectures, \sys can generate binaries for each arch. We defer this to future work.

\noindent{\bf Can we achieve better profiling?}
In \sys, resource efficiency heavily relies on profiling accuracy (\cref{subsub:pipe_para} and~\cref{subsec:app_orch}). 
Our current approach uses offline profiling, yet this may not accurately reflect runtime performance, especially for changing traffic patterns. 
Luckily, \sys supports adaptive scaling at runtime, and we can still conduct profiling even after application deployment. 
When a gap between offline profiling and runtime results exceeds a predefined threshold,
resource allocator and adaptive scaling can be re-applied,  enhancing overall resource efficiency. 
The runtime profiling can be periodically conducted with adjustable time intervals.

\section{Related Work}
\label{sec:related}

We now discuss the related prior efforts in addition to \cite{ns21,db23,dyk23,le17uno} in \cref{sec:introduction}.

One line of existing work focuses on co-offloading applications' logic on the same \nic to reap the multiplexing potential \cite{panic20,supernic22,nica19,fairnic20}. 
They either enforce isolation abstractions on existing \snics \cite{nica19,fairnic20} 
or design new \snics with fair resource sharing \cite{panic20,supernic22}.
Vendors are also exploring multi-host \nics that integrate multiple PCIe interfaces, each connecting to an independent host \cite{multihost1,multihost2,multihost3}. 
While they prove the multiplexing benefits on \snics, they share the local offloading assumption without a global view of the cluster, resulting in similar management inefficiencies discussed in \cref{subsec:use}.

Prior research has also explored microservice-based \snic orchestration \cite{e319},
enhancing microservices' energy efficiency by offloading with low-power \nic resources.
Building upon existing host microservice platforms \cite{sf19}, this approach introduces heterogeneous communication and microservice orchestration between hosts and \nics.
However, it inherits the non-trivial communication overhead from the microservice's RPC stack \cite{ganmic,nikita21dagger}, 
and deploys entire microservices to monolithic \nics, scaling onboard resources at the per-\nic granularity, leading to suboptimal resource efficiency. 
In contrast, \sys is a unified platform for common \nic applications favoring simpler traffic redirection, and introduces finer-grained resource scaling across the cluster. 
This requires novel designs on the programming model, data plane, and control plane, which remain untouched in prior work.

\section{Conclusion}
\label{sec:conclusion}

We proposed a new vision of \snic as a service to address the inefficiencies in application deployment and resource management of \snic clusters. 
We showcased our system, \sys, that realizes this vision by efficiently pooling \nic resources and exposing a one-\nic abstraction to application developers. 
It incorporates a parallelizable data plane for scalable performance and fine-grained resource scaling to enhance overall efficiency. 
Our extensive testbed experiments on heterogeneous commodity \snics demonstrate that \sys improves resource efficiency by up to 1.75$\times$ compared to 
state-of-the-art systems, and delivers scalable throughput with small latency overheads.

\balance
\bibliographystyle{plain}
\bibliography{main}

\clearpage
\nobalance

\appendix
\section*{Appendices}

\section{\sys APIs}
\label{subsec:meili_api_list}

The partial list of \sys's APIs is shown in Table~\ref{table:sys_api}. \texttt{UCF} denotes the user-customized function.

\begin{table}[h]
    \centering
    \resizebox{\columnwidth}{!}{
    \begin{tabular}{ll}
    \toprule
    {\bf Function} & {\bf Description} \\ \midrule 
    {\tt pkt\_trans({\tt UCF, pkt})} & Run a packet transformation operation.\\ 
    {\tt pkt\_flt({\tt UCF, pkt})}  & Filter packets with the operation. \\  
    {\tt flow\_ext({\tt UCF, wnd, slide, stream})}  & Construct flows from a stream. \\ 
    {\tt flow\_trans({\tt UCF, flow})}  & Run a flow transformation operation. \\ 
    {\tt reg\_sock({\tt app\_name, sock\_id})}  & Register an established socket to \sys. \\ 
    {\tt epoll({\tt UCF, sock\_id, event})}  & Process an event on the socket. \\ 
    {\tt Regex({\tt addr, rules})}  & The built-in Regular Expression API. \\ 
    {\tt AES({\tt addr, key})}  & The built-in AES Encryption API. \\ 
    {\tt Compress({\tt addr, rt})}  & The built-in Compression API. \\ 
    {\tt s\_add({\tt name, state})}  & Add the state. \\ 
    {\tt s\_remove({\tt name, state})}  & Remove the state. \\ 
    {\tt s\_set({\tt name, state})}  & Set the state . \\ 
    {\tt s\_get({\tt name, state\_idx})}  & Find and get one state. \\ 
    {\tt s\_traverse({\tt state\_name})}  & Traverse the states. \\ 
    {\tt s\_compute({\tt state\_name, UCF})}  & Enforce the state computation. \\ 
    {\tt app\_sub\_thr({\tt app, target})}  & Submit app with a Tput target.\\ 
    \bottomrule
    \end{tabular}
    }
    \caption{A partial list of \sys's APIs.}
    \vspace{-5mm}
    \label{table:sys_api}
\end{table}

\section{Socket Processing Example}
\label{subsec:sock_ex}

The pseudocode of a socket processing application is shown below. 
The application implements an API gateway \cite{e319}.
\texttt{epoll\_in()} is a \uco and submitted via \sys's \texttt{epoll()} operation.

\begin{lstlisting}
    // User-customized functions
    void epoll_in(sid, event) {
        if (event != EPOLLIN) exit(-1);
        ret = read(sid, buf, SIZE);
        // Authenticate the API call
        hmac_recv = buf.hmac;
        hmac = sha(buf, BLK_SIZE);
        if (hmac_recv == hmac)
            rate_limit(buf);
            redirect(backend);
    }
    // Meili API invocation
    Meili.reg_sock(sid);
    Meili.epoll(epoll_in, sid, EPOLLIN);
\end{lstlisting}

\section{Failover}
\label{subsec:failover}

To ensure \snic application availability, \sys provides crucial failure detection and recovery mechanisms. 
It considers the failure domain where the \snic fails or becomes unreachable due to its network link failure, 
and leverages the failover manager in existing platforms \cite{k8s,azurefunc,awslam} and extends the failure detection and recovery.
This involves periodic replication of flow states and packet caching to a backup \nic, serving as a failover replica in case of primary \nic failure. 
\sys Controller periodically checks primary \nic availability via connection requests. 
Upon detecting a failure, it promptly launches a new (sub-)pipeline on the replica, recovering from the latest synchronized states and processing cached packets. 
\sys also configures traffic redirection policies on \tor to guide incoming traffic to the replica.

\begin{figure}[t]
    \centering
    \includegraphics[width=0.95\linewidth]{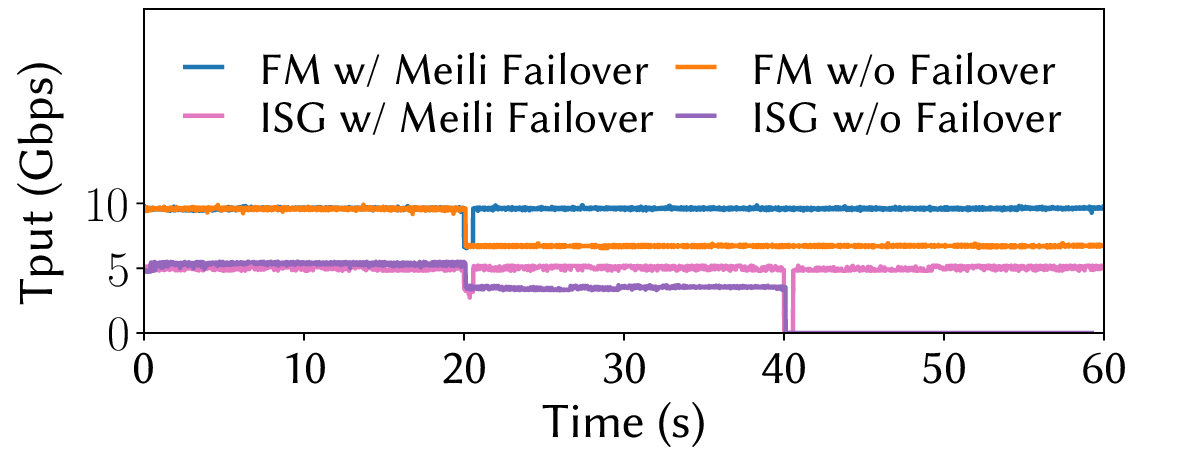}
    \vspace{-5mm}
    \caption{Throughput of two applications with and without \sys's failover.}
    \label{fig:failover}
\end{figure}

\noindent{\bf Evaluation.}
We examine \sys's failure recovery using two BlueField-2 (\texttt{NIC1} and \texttt{NIC2}) 
and one Pensando \snic (\texttt{NIC3}) to deploy a Flow Monitor (FM) and an IPsec Gateway (ISG), 
with another BlueField-2 and Pensando \snic for failover replica.
Flow Monitor achieves a 10Gbps target utilizing CPU cores on \texttt{NIC1} and \texttt{NIC2}, 
while IPsec Gateway uses CPU cores on these two \snics\ and AES accelerator on \texttt{NIC3} to reach a 5Gbps target. 
We generate 32 flows with 1500B packet sizes for each application via DPDK-Pktgen. 
The failure is emulated by disabling \snic's network ports.
Figure~\ref{fig:failover} shows the throughput during \texttt{NIC2} failure at 20s and \texttt{NIC3} failure at 40s. 
At 20s, Flow Monitor and IPsec Gateway have a throughput drop, with pipelines on \texttt{NIC1} and \texttt{NIC3} remaining active. 
However, at 40s, when \texttt{NIC3} fails, IPsec Gateway becomes inaccessible due to AES accelerator's unavailability, impacting its entire pipeline. 
We see the throughput recovers within 500ms, proving \sys's ability to maintain application availability.

\end{document}